\documentclass{article}
\usepackage[title]{appendix}
\usepackage{graphicx} 
\usepackage{fullpage,times,fourier,charter}
\usepackage{caption}
\usepackage{subcaption}
\usepackage{authblk}
\usepackage{graphicx,xcolor}
\usepackage[title]{appendix}
\usepackage{fullpage,fourier,charter}
\usepackage{amsmath}
\usepackage{mathtools}  
\usepackage{verbatim}
\usepackage{xfrac}
\usepackage{amsthm,amssymb,bm}
\usepackage[colorlinks=true,citecolor=blue]{hyperref}

\def\arc{\mathop{\rm arc}\nolimits}
\def\sech{\mathop{\rm sech}\nolimits}

\numberwithin{equation}{section}
\advance\textheight 10mm
\advance\voffset -2mm
\def\bse{\begin{subequations}}
\def\ese{\end{subequations}}

\makeatletter

\renewcommand\section{\@startsection {section}{1}{\z@}%
  {-2.6ex \@plus -1ex \@minus -.2ex}{1.4ex \@plus.1ex}%
  {\large\bf\sffamily}}
\renewcommand\subsection{\@startsection{subsection}{2}{\z@}%
  {-2.2ex\@plus -0.4ex \@minus -.2ex}{1.2ex \@plus .1ex}%
  {\normalfont\small\bf\sffamily}}
\renewcommand\subsubsection{\@startsection{subsubsection}{3}{\z@}%
  {-1ex\@plus -0.2ex \@minus -.2ex}{0.4ex \@plus .1ex}%
  {\normalfont\normalsize\it}}
\makeatother

\title{First-order continuum models for nonlinear dispersive waves in the granular crystal lattice}
\author[1]{Su Yang}
\author[2]{Gino Biondini}
\author[3]{Christopher Chong}
\author[1]{Panayotis G. Kevrekidis}
\affil[1]{\small\it Department of Mathematics and Statistics, University of Massachusetts, Amherst, 01003-4515, Massachusetts, USA}
\affil[2]{\small\it Department of Mathematics, University at Buffalo, Buffalo, NY 14260-2900, USA}
\affil[3]{\small\it Department of Mathematics, Bowdoin College, Brunswick, ME 04011, USA}
\date{\small\today}

\begin{document}

\maketitle
\begin{abstract} 
    We derive and analyze, analytically and numerically, two first-order continuum models to approximate the nonlinear dynamics of granular crystal lattices, focusing specifically on solitary waves, periodic waves, and dispersive shock waves. The dispersive shock waves predicted by the two continuum models are studied using modulation theory, DSW fitting techniques, and direct numerical simulations. The PDE-based predictions show good agreement with the DSWs generated by the discrete
    model simulation of the granular lattice itself, even in cases where no precompression is present and the lattice is purely nonlinear. Such an effective
  description could prove useful for future, more analytically amenable
    approximations of the original lattice system.
\end{abstract}

\tableofcontents

\section{Introduction}

Over the past few decades, the realm of granular 
crystals has offered a fertile playground for
the exploration of nonlinear wave phenomena.
Relevant advances have, by now, been summarized
in a wide range of reviews, as well as books~\cite{Nester2001,granularBook,yuli_book,gc_review,sen08}.
The relevant developments concern a wide range of
coherent structures, including traveling 
waves~\cite{Nester2001,sen08}, discrete 
breathers~\cite{granularBook,yuli_book,gc_review},
as well as more recently the realm of dispersive
shock waves (DSWs)~\cite{chong2024dynamicsnonlinearlattices}.

The study of DSWs, more specifically, has been 
a subject gaining considerable traction, not only
in the setting of mechanical metamaterials, but also in
other areas blending dispersive and nonlinear features,
including but not limited to superfluids and atomic
gases, nonlinear optics, water waves, and plasmas,
as summarized, e.g., in~\cite{scholar,Mark2016,Whitham74}. 
Interestingly, the granular realm has offered
some of the early experimental realizations of {\it discrete}
DSWs, e.g., in the works of~\cite{Herbold07} 
and~\cite{Molinari2009} (the latter in a dimer setting),
at around the same time as relevant experiments materialized
in optical waveguide arrays~\cite{fleischer2}. More
recently, such coherent structures have emerged
in chains of hollow elliptic cylinders~\cite{HEC_DSW},
as well as in tunable magnetic lattices~\cite{talcohen}.
These more recent examples have enabled systematic
visualization capabilities of the space-time evolution
of DSWs through experimental techniques such as laser 
Doppler vibrometry.

Some of these experimental 
works have attempted to make partial
connections with the corresponding theoretical 
expectations.  
For instance, the work of~\cite{Molinari2009}
sought to characterize the traveling waves in a
dimer lattice at the front of the observed DSW.
Nevertheless, the theoretical approach based on the
Whitham modulation theory that has been developed
for lattice systems, e.g., in~\cite{Venakides99,DHM06,blochkodama}
does not seem to have
caught on with respect to DSW detailed computations
and associated experimental observations.
In light of that, the present authors in a series of
works have sought to develop tools either 
leveraging integrable models such as the Korteweg-de Vries
(KdV) equation and the Toda lattice~\cite{Ari2024,physd2024v469p134315}, 
or by adapting asymptotic techniques such
as the DSW fitting method of~\cite{El_2005} to quantitatively
characterize
problems in such discrete~\cite{Sprenger2024} 
or/and metamaterial settings~\cite{yang2024regularizedcontinuummodeltraveling}.

The present work constitutes a significant further
step in this program. Indeed, one of the most
canonical connections that exists for Fermi-Pasta-Ulam-Tsingou (FPUT)
lattices~\cite{FPU55,Berman2005,FPUreview,zabu}
is that of the KdV equation. The latter has been
used not only to rigorously approximate the results
of FPUT, but also to characterize the wave stability
and dynamics therein~\cite{pegogf1,pegof2,pegof3,pegof4}.
Nevertheless, as is well-known, such a unidirectional
model (like the KdV) is crucially obtained when
the granular problem possesses linear dispersion,
e.g., in the form of the so-called precompression~\cite{Nester2001,sen08,granularBook}.
In the absence of precompression, we are not
aware of a {\it well-posed}, unidirectional model
that is capable of capturing the dynamics 
in the so-called sonic vacuum~\cite{Nester2001} regime,
with the notable exception of~\cite{pelingj} which, however,
operates in the vicinity of the linear limit of the
exponent $p=1$ (see below).

Motivated by this feature, we propose a novel (to the
best of our knowledge) model corresponding to a generalized
KdV equation which is applicable both in the presence
but also in the absence of precompression. 
We also regularize the relevant model, in a way
analogous to how this is done for the KdV to obtain
so-called Benjamin-Bona-Mahony (BBM) equation~\cite{Benjamin1972}.
For both of these proposed models, with an aim towards
examining the DSW phenomenology, we analyze their
solitary wave solutions first and subsequently
their periodic waveforms, as well as the corresponding
conservation laws. Upon detailing these, we examine
the Whitham modulation system and consider
the models' rarefaction wave. We also provide
the DSW fitting as a way to examine the properties
of both the leading and the trailing edge (namely,
the amplitude and speed of the former, as well as the
speed and wavenumber of the latter). Equipped with these
theoretical tools, we compare the results of the fully
discrete granular chain with those of the KdV for finite
precompression, as well as with those of our newly proposed models both with and without precompression.
Interestingly, we find that the newly proposed models feature demonstrably better predictions than the KdV (i.e., more proximal to the discrete model) for finite but small precompression, while the results of the different models become comparable to the KdV as the precompression is enhanced.

\section{First-order continuum models}

\subsection{The granular crystal lattice and first-order in time continuum models}
\label{s:models}

In this work, we focus on the granular crystal lattice system whose (normalized) equations of motion can be described by the following differential-difference equations,
\begin{equation}
\label{e:granular lattice on displacement}
    \ddot u_n = \left[\delta_0 +u_{n-1}-u_{n} \right]_+^{p} - \left[\delta_0+u_n - u_{n-1}\right]_+^{p},
\end{equation}
where $u_n$ refers to the displacement of the $n$th bead in the granular chain and $\delta_0$ represents a static precompression. 
When two adjacent beads come out of contact, there is no force, which is captured by the ``rectification'' operator $\left[f\right]_+ = \max(f,0)$.

For the present paper, it will be more convenient to work with the strain variables $r_n = u_{n-1} - u_{n}$, in which case the equations of motion becomes
\begin{equation}
\label{Granular Crystals}
    \ddot r_n = \left(r_{n+1}\right)^{p} - 2\left(r_n\right)^{p} + \left(r_{n-1}\right)^{p},
\end{equation}
where we have taken $\delta_0 = 0$. 
In what follows, we will always assume that $r_n\geq0$, in which case we can drop the rectification operator.
Looking for a plane-wave solution of the form 
$r_n(t) = A + Be^{i\left(kn - \omega t\right)}$ 
of~\eqref{Granular Crystals}, where $\left|B/A\right| \ll 1$, yields the linearized dispersion relation
\begin{equation}
\label{e:linearized DR of DDE}
    \omega^2 = 4pA^{p-1}\sin^2\left(\frac{k}2\right).
\end{equation}

\begin{figure}[t!]
    \centering
    \includegraphics[width=0.875\linewidth]{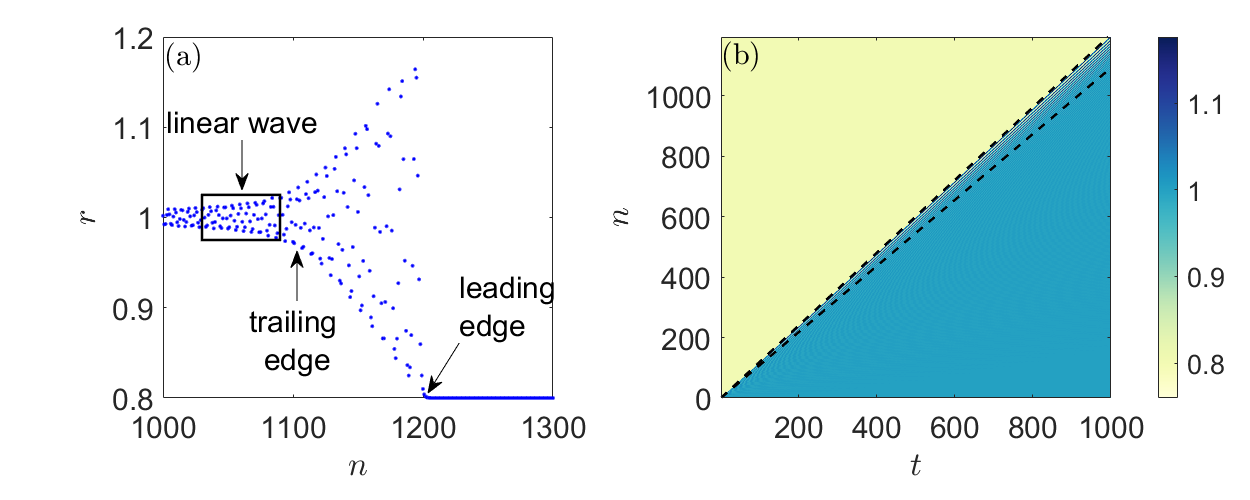}
    \caption{
    \textbf{(a)}  Numerical simulations of the Riemann problem~\eqref{Granular Crystals} with $r^+ = 0.8$ and $r^- = 1$.  
    (Please see section~\ref{s:numerics} for a discussion of the precise ICs used in the numerical simulations.)
    The spatial profile of the granular lattice DSW with $p = 3/2$ is shown at $t = 1000$. 
    \textbf{(b)} Density plot of the DSW corresponding to the left panel.
    The two dashed black lines represent, respectively, the leading (upper) and the trailing (lower) edge of the DSW, and are given by the expressions $n = s^+t$ (upper) and $n = s^-t$ (lower), with $s^\pm$ defined later in the text 
    [see \eqref{e:non-reg trailing-edge speed} and \eqref{e:non-reg leading-edge speed}].}
    \label{fig:DSW introduction}
\end{figure}

The main focus of this paper is the study of dispersive shock wave phenomena for the granular lattice described by~\eqref{Granular Crystals}, which can be produced by the following so-called Riemann initial data,
\begin{equation}\label{e:Riemann IC for the granular lattice}
    r_n(0) = \begin{cases}
        r^-, \hspace{2mm} n \leq 0,\\
        r^+ ,\hspace{2mm} n > 0,
    \end{cases}
    \hspace{5mm}
    \dot r_n(0) = \begin{cases}
        v^-, \hspace{2mm} n \leq 0,\\
        v^+, \hspace{2mm} n > 0.
    \end{cases}
\end{equation}
An example DSW that results from Riemann initial data is shown in Fig.~\ref{fig:DSW introduction}. In this figure, key features of the DSW, such as the trailing (linear) and leading (solitonic) edges can be seen. These features, as well as the DSW profile itself
will be analyzed using various PDE models, which we detail next.

The goal of this work is to derive and analyze appropriate continuum 
models to approximate the discrete granular chain model \eqref{Granular Crystals}. 
The main idea is to use a dispersive, long-wavelength model to that effect by introducing the following two slowly varying spatial and temporal scales:
\begin{equation}
\label{e:slowly varying scales}
    X = \epsilon n, \hspace{3mm} T = \epsilon t, 
\end{equation}
where $0 < \epsilon \ll 1$ is a smallness parameter. Then, substituting \eqref{e:slowly varying scales} into Eq.~\eqref{Granular Crystals} yields, 
\begin{equation}
\label{e:substitution of the ansatz}
    \epsilon^2 r_{TT} = r^{p}\left(X+\epsilon,T\right) - 2r^{p}\left(X,T\right) + r^{p}\left(X-\epsilon,T\right).
\end{equation}
A Taylor expansion of~\eqref{e:substitution of the ansatz} yields, to leading order, 
\begin{equation}
\label{e:Taylor expansion leading-order}
    r_{TT} = \left(r^{p}\right)_{XX}.
\end{equation}
However, the PDE \eqref{e:Taylor expansion leading-order} is dispersionless 
which can be readily seen since its linearized dispersion relation has zero second derivative with respect to the wave number. 
This indicates that it cannot be used as a model to capture the dispersive shock wave emerging from the discrete model~\eqref{Granular Crystals}. 
To resolve this issue, one natural strategy is to include the higher-order terms in the expansion of \eqref{e:substitution of the ansatz}. 
Specifically, keeping the terms with order of $\mathcal{O}\left(\epsilon^2\right)$, one obtains 
\begin{equation}
\label{e:higher order terms in Eq}
    r_{TT} = \left(r^{p}\right)_{XX} + \frac{\epsilon^2}{12}\left(r^{p}\right)_{XXXX}.
\end{equation}
Now, looking for a plane-wave solution in the form of $r\left(X,T\right) = A + Be^{i\left(KX-\Omega T\right)}$ where $\left|B/A\right| \ll 1$ for the PDE \eqref{e:higher order terms in Eq} yields the linearized dispersion relation
\begin{equation}
\label{e:LDR for the higher order model}
    \Omega^2 = pA^{p-1}K^2\left(1 - \frac{1}{12}\epsilon^2K^2\right).
\end{equation}
Note that the value of $\Omega$ is purely imaginary for sufficiently large $K$ 
(corresponding to modulational instability).  More importantly, the imaginary part of $\Omega$ is unbounded for large~$K$.
This suggests that~\eqref{e:LDR for the higher order model} is ill-posed and therefore not a good model to approximate the DSW of the granular chain. 
A second-order in time regularization of the model~\eqref{e:higher order terms in Eq} was recently derived and studied in \cite{yang2024regularizedcontinuummodeltraveling}, and was shown to provide a good description 
of the DSW of~\eqref{Granular Crystals}.

\begin{figure}[t!]
    \centering
    \includegraphics[width=0.8\linewidth]{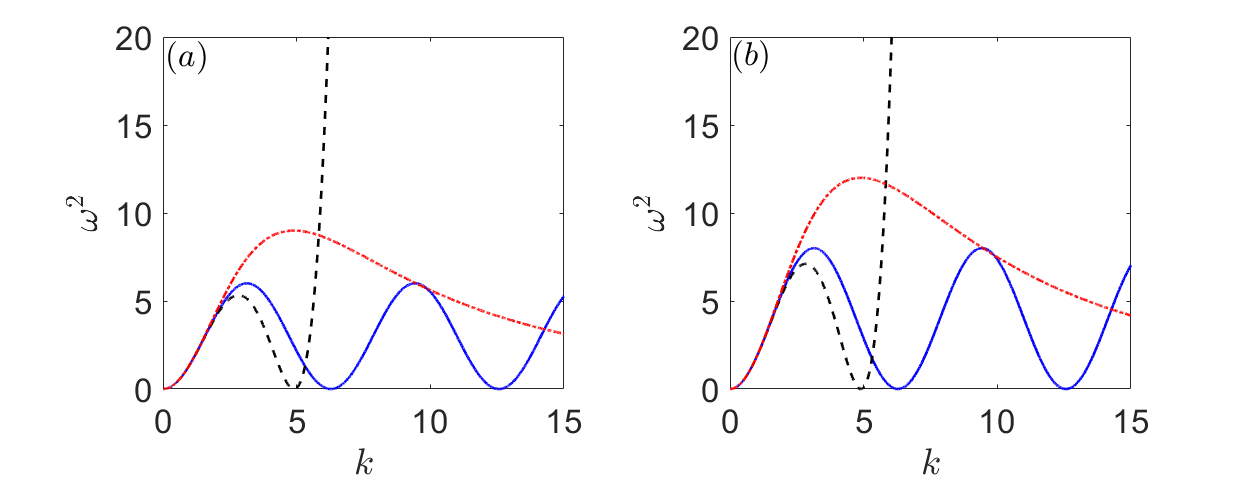}
    \caption{Comparison of linearized dispersion relations. 
    The solid blue, dashed black, and dotted dashed red curves depict the linearized dispersion relations of~\eqref{Granular Crystals}, \eqref{e:KdV-like PDE} and \eqref{e:BBM-like PDE}, respectively. 
    The values of relevant parameters are: \textbf{(a)} $p = 3/2$ and $A = 1$; \textbf{(b)} $p = 2$ and $A = 1$.
    }
    \label{fig1: linearized DR}
\end{figure}

A different way to obtain a well-posed model is to consider the positive branch of the dispersion relation~\eqref{e:LDR for the higher order model} (corresponding to right-going waves), namely
\begin{equation}
\label{e:Positive sign of LDR}
    \Omega = \sqrt{p}A^{\frac{p-1}2}K\sqrt{1 - \frac{1}{12}\epsilon^2K^2},
\end{equation}
Taking the long-wave limit ($0 < K \ll 1$) of the linear dispersion relation by Taylor expanding~\eqref{e:Positive sign of LDR} near $K = 0$, we obtain
\begin{equation}
\label{e:final LDR}
    \Omega(K,A) = \sqrt{p}A^{\frac{p-1}2}K\left(1 - \frac{1}{24}\epsilon^2K^2\right).
\end{equation}
We then observe that a (non-regularized) continuum model associated with the linear dispersion relation~\eqref{e:final LDR} reads
\begin{equation}
    r_T + \frac{2\sqrt{p}}{p+1}\left(r^{\frac{p+1}2}\right)_X + \frac{\sqrt{p}\epsilon^2}{12\left(p+1\right)}\left(r^{\frac{p+1}2}\right)_{XXX} = 0.
\label{e:KdV-like PDE}
\end{equation}
Equation~\eqref{e:KdV-like PDE} will serve as a first continuum approximation for the discrete model~\eqref{Granular Crystals}.
Moreover, we also note that~\eqref{e:KdV-like PDE} can be regularized so as to be lower order in the spatial derivatives and, correspondingly, to have a bounded dispersion relation as $K\to\infty$. 
To this end, we first rewrite~\eqref{e:KdV-like PDE} as
\begin{equation}
\label{e:KdV_like in operator form}
    r_T = -\frac{2\sqrt{p}}{p + 1}\partial_X\left(1+\frac{\epsilon^2}{24}\partial^2_{X}\right)r^{\frac{p+1}2}.
\end{equation}
Inverting the operator $1+\frac{\epsilon^2}{24}\partial^2_{X}$ on the RHS of~\eqref{e:KdV_like in operator form} then yields the following regularized continuum model:
\begin{equation}
\label{e:BBM-like PDE}
    r_T - \frac{\epsilon^2}{24}r_{XXT} = -\frac{2\sqrt{p}}{p + 1}\left(r^{\frac{p+1}2}\right)_X.
\end{equation}
Note that if $p=2$, then Eq.~\eqref{e:BBM-like PDE} is the Benjamin–Bona–Mahony
equation \cite{Benjamin1972}.
Looking for a plane wave solution of~\eqref{e:BBM-like PDE} yields the following linear dispersion relation
\begin{equation}
\label{eq:LDR for BBM}
    \Omega(K,A) = \frac{\sqrt{p}A^{\frac{p-1}2}K}{1 + \frac{\epsilon^2K^2}{24}}.
\end{equation}
A comparison of the three dispersion relations in~\eqref{e:linearized DR of DDE}, \eqref{e:final LDR} and \eqref{eq:LDR for BBM} is displayed in Figure~\ref{fig1: linearized DR}.

The two continuum models in~\eqref{e:KdV-like PDE} and \eqref{e:BBM-like PDE} will be the main focus of this paper, and will be utilized to approximate both the solitary waves and the DSWs of the granular chain \eqref{Granular Crystals}. 
Before we step into the actual details of these two models, however, we first briefly turn our attention to another continuum model, the Korteweg–de Vries (KdV) equation. 
This is because the DSWs of the KdV equation are very well studied
and were also recently used in \cite{Chong_2024} to approximate those of the granular crystal lattice~\eqref{Granular Crystals}. 
The KdV approximation will therefore serve as a benchmark for our study.

\subsection{KdV approximation and its limitations}
\label{s:KdV approx}

In this section, we briefly review the main features of the KdV approximation,
in view of a comparison with the novel models introduced in section~\ref{s:models}. 
To this end, we first perform the following change of variable on the strain:
\begin{equation}
\label{e:Change of variable of r_n}
    y_{n} = r_n - r^+,
\end{equation}
so that the equation of motion \eqref{Granular Crystals} now becomes
\begin{equation}
\label{e:eqn for y_n}
    \ddot y_n = (r^++y_{n-1})^{p} - 2(r^++y_n)^{p}+(r^++y_{n+1})^{p},
\end{equation}
with initial conditions
\begin{equation}
    y_n(0) = \begin{cases}
        r^--r^+, \quad & n\leq0,\\
        0, \quad & n >0.
    \end{cases}
\end{equation}
Below we reduce \eqref{e:eqn for y_n} to a first-order in time system, 
and comment on how to initialize the second order and first order
problems so that they are ``consonant'' with each other.
A Taylor expansion of~\eqref{e:eqn for y_n} leads to the following FPUT equation:
\begin{equation}
\label{e:y_n FPUT eqn}
    \ddot y_n = K_2(y_{n-1}-2y_n+y_{n+1}) + K_3(y_{n-1}^2-2y_{n}^2+y_{n+1}^2),
\end{equation}
where $K_2 = p(r^+)^{p-1}$ and $K_3 = p(p-1)(r^+)^{p-2}/2$.
Similarly to the previous section, we then perform the following change of variables:
\begin{equation}
    y_n = \epsilon^2Y(X,T), \quad X = \epsilon(n-\sigma t), \quad T = \epsilon^{3}t,
\end{equation}
where $0<\epsilon\ll1$ is a formal smallness parameter and $\sigma = \sqrt{K_2}$ is the sound speed.
Then, the KdV reduction of the granular chain can be obtained by collecting terms at the order of $\mathcal{O}(\epsilon^{6})$, which reads,
\begin{equation}\label{e:KdV reduction}
    Y_{T} + \frac{K_3}{\sigma}YY_{X} + \frac{\sigma}{24}Y_{XXX} = 0.
\end{equation}
Notice that the consistent initial data for the KdV equation \eqref{e:KdV reduction} reads
\begin{equation}
    Y(X,0) = \begin{cases}
        \epsilon^{-2}(r^- - r^+), \quad & X \leq 0,\\
        0, \quad & X > 0.
    \end{cases}
\end{equation}
Furthermore, to simplify our analysis of the DSW, we transform~\eqref{e:KdV reduction} into the following standard form of the KdV equation:
\begin{equation}\label{e:KdV reduction in unit form}
   \widetilde{Y}_\tau + \widetilde{Y}\widetilde{Y}_X + \widetilde{Y}_{XXX} = 0,
\end{equation}
where 
\begin{equation}
    \widetilde{Y}(X,\tau) = \frac{24K_3}{\sigma^2}Y(X,T), \quad \tau = \frac{\sigma T}{24}.
\end{equation}
with the initial condition
\begin{equation}\label{e:Initial data for Y_tilde}
    \widetilde{Y}(X,0) = \begin{cases}
        {24K_3(r^--r^+)}/{(\sigma^2\epsilon^2)}, \quad & X \leq 0,\\
        0, \quad & X > 0.
    \end{cases}
\end{equation}

We now recall the theoretical prediction of the edge quantities of the KdV DSW from \cite{GP73,El1995}. In particular, the solitonic-edge amplitude $a^+_{\text{KdV}}$, speed $s^+_{\text{KdV}}$, linear-edge speed $s^-_{\text{KdV}}$, and wavenumber $k^-_{\text{KdV}}$ are given as follows:
\begin{equation}
\label{e:KdV  DSW theoretical predictions}
    a^+_{\text{KdV}} = 2\Delta, \quad s^+_{\text{KdV}} = \frac2{3}\Delta,\qquad
    s^-_{\text{KdV}} = -\Delta, \quad K^-_{\text{KdV}} = \sqrt{\frac{2\Delta}{3}},
\end{equation}
where $\Delta = {24K_3(r^--r^+)}/({\sigma^2\epsilon^2})$ denotes the initial jump. 
Moreover, to compare these theoretical predictions with the numerically measured DSW edge features of the granular lattice \eqref{Granular Crystals}, we need to rescale the results in~\eqref{e:KdV  DSW theoretical predictions} as follows:
\begin{equation}
\label{e:KdV prediction in the granular lattice}
    a^+ = \frac{\epsilon^2\sigma^2}{24K_3}a^+_{\text{KdV}},\quad s^+ = \frac{\sigma\epsilon^2}{24}s^+_{\text{KdV}} + \sigma,\qquad
    s^- = \frac{\sigma\epsilon^2}{24}s^-_{\text{KdV}}+\sigma,\quad k^- = \epsilon K^-_{\text{KdV}}.
\end{equation}

\begin{figure}[t!]
    \centering
    \includegraphics[width=0.75\linewidth]{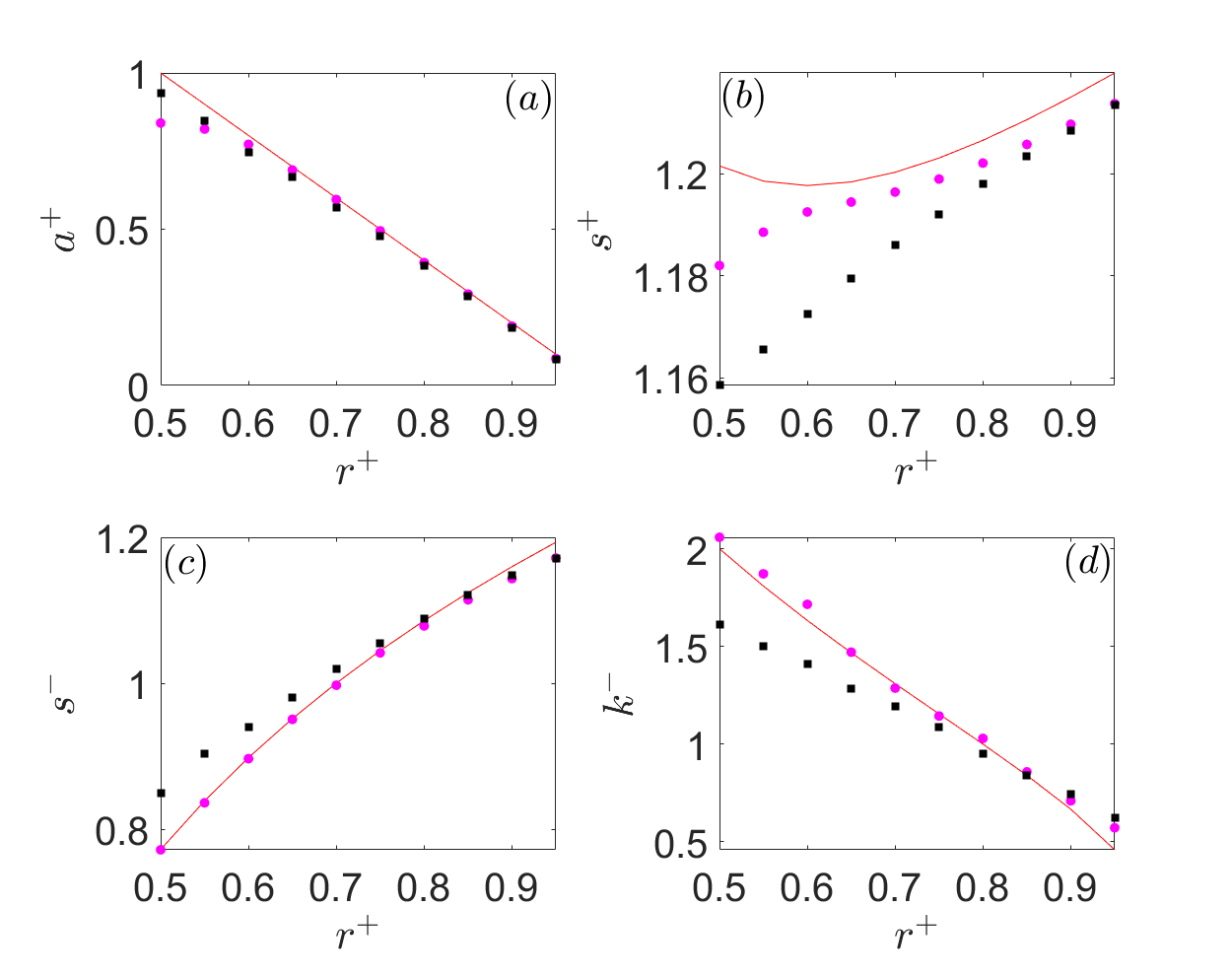}
    \caption{Comparison of the KdV theoretical predicted DSW-edge features with the numerically measured DSW-edge features of the granular lattice \eqref{Granular Crystals}. \textbf{(a)} The leading-edge amplitude $a^+$, \textbf{(b)} The leading-edge speed $s^+$, \textbf{(c)} The trailing-edge speed $s^-$, and \textbf{(d)} The trailing-edge wavenumber $k^-$. Notice that the red curve in each panel depicts the KdV prediction based on Eq.~\eqref{e:KdV prediction in the granular lattice}, while the magenta circles and the black squares refer to the numerically measured DSW-edge features of the KdV reduction Eq.~\eqref{e:KdV reduction} and the granular lattice \eqref{Granular Crystals}, respectively.
    }
    \label{fig: KdV reduction DSW edge features}
\end{figure}

Next, we compare the theoretical predicted KdV DSW-edge features in~\eqref{e:KdV prediction in the granular lattice} with the numerically computed DSW-edge features of the continuum KdV reduction in~\eqref{e:KdV reduction} and those of the granular lattice \eqref{Granular Crystals}. 
Since the granular lattice \eqref{Granular Crystals} is a second-order system, we need appropriate initial data for the velocity $\dot r_n(0)$. This initial velocity is determined by the KdV equation~\eqref{e:KdV reduction}
as follows:
\begin{equation}
    \dot r_n(0) = \epsilon^{5}\bigg(-\frac{K_3}{\sigma}Y(X,0)Y_{X}(X,0) -\frac{\sigma}{24}Y_{XXX}(X,0)\bigg) - \epsilon^{3}\sigma Y_{X}(X,0).
\end{equation}
As a relevant note, $\epsilon = 0.1$ is used throughout the whole paper, and we apply a spectral integrating factor in space and an exponential time differencing RK$4$ (ETDRK$4$) scheme in time stepping \cite{doi:10.1137/S1064827502410633} to simulate the KdV equation. 
 
\begin{figure}[t!]
    \centering
    \includegraphics[width=1.05\linewidth]{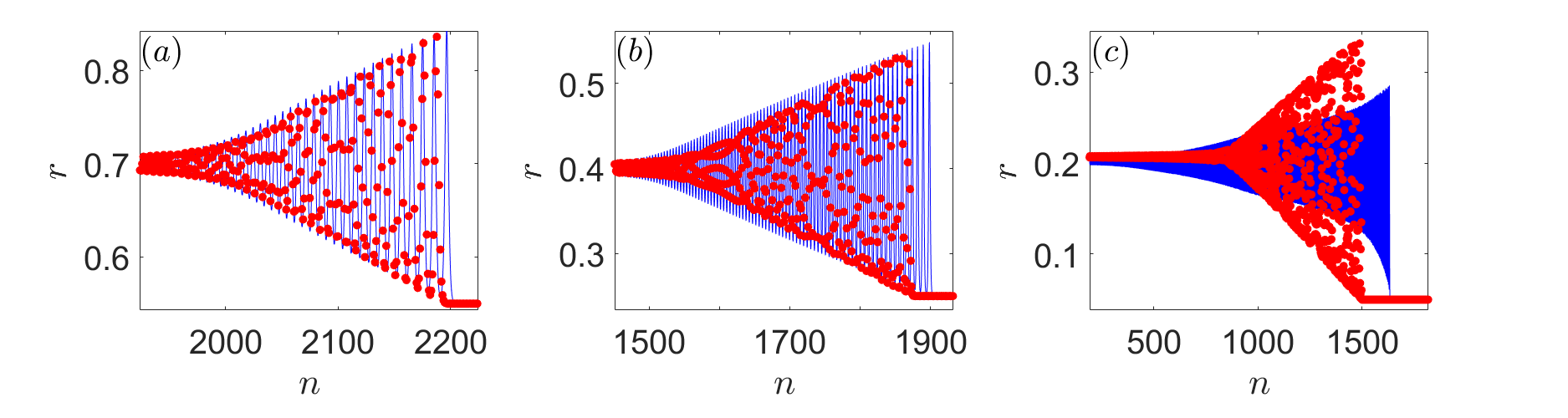}
    \caption{
    Comparison between the spatial profile of the KdV DSW with that of the granular chain at $t = 2000$ for $p = 3/2$.  
    In all three panels, the blue solid curves display the DSW of the KdV equation \eqref{e:KdV reduction}, while the discrete red circles showcase the DSW of the granular chain. Furthermore, $r^+ = 0.55, 0.25, 0.05$ from the leftmost to the rightmost panel, 
    respectively, where the jump between $r^+$ and $r^-$ is fixed to be $\Delta=0.15$.
    }
    \label{fig:KdV DSW spatial profile comparisons}
\end{figure}

Figure~\ref{fig: KdV reduction DSW edge features} showcases the comparison of the KdV theoretical predictions in~\eqref{e:KdV prediction in the granular lattice} with the associated numerically measured DSW-edge features of the continuum KdV reduction in~\eqref{e:KdV reduction} and those of the discrete granular chain \eqref{Granular Crystals}
(see Appendix for details on the numerical calculation of the trailing edge and leading edge speeds).
We observe that as the value of lower background $r^+$ decreases, the KdV reduction performs worse in approximating the DSW of the granular chain, as expected. This can also be seen by inspection of the spatial profile comparison of the KdV-reduction DSW and the granular DSW, see Fig.~\ref{fig:KdV DSW spatial profile comparisons}.  
More importantly, however, the KdV prediction is only valid in the presence of precompression (or equivalently for $r^+>0$). 
Below we will demonstrate that the proposed continuum models~\eqref{e:KdV-like PDE} and~\eqref{e:BBM-like PDE} not only perform better than the KdV approximation,
but they are also valid in the absence of precompression, namely for a purely nonlinear chain.

\section{Solitary wave and periodic traveling wave solutions}

Having motivated the need for continuum approximations beyond the classical KdV one in the previous section, we now begin
our investigation of the two continuum models proposed in section~\ref{s:models}.
We start with the traveling wave solutions to both models \eqref{e:KdV-like PDE} and \eqref{e:BBM-like PDE}.

\subsection{Solitary wave solutions}
\label{s:Soliton solutions}

\paragraph{Solitary waves in the non-regularized model.}

For the non-regularized model \eqref{e:KdV-like PDE} 
we first take the following traveling-wave ansatz,
\begin{equation}
    r\left(X,T\right) = R(Z), \hspace{5mm} Z = X-cT,
    \label{Traveling-wave ansatz}
\end{equation}
where $c \in \mathbb{R}$ denotes the propagation speed of the solitary wave.
Substituting the ansatz \eqref{Traveling-wave ansatz} into the continuum PDE \eqref{e:KdV-like PDE} yields
\begin{equation}
    -cR_Z + \frac{2\sqrt{p}}{p+1}\left(R^{\frac{p+1}2}\right)_Z + \frac{\sqrt{p}\epsilon^2}{12\left(p+1\right)}\left(R^{\frac{p+1}2}\right)_{ZZZ} = 0.
    \label{Traveling-wave ODE}
\end{equation}
We then apply the change of variables $v = R^{\frac{p+1}2}$, and integrate the ODE \eqref{Traveling-wave ODE} twice to obtain that
\begin{equation}
    \frac{\sqrt{p}\epsilon^2}{24\left(p+1\right)}\left(v_Z\right)^2 = -\frac{\sqrt{p}}{p+1}v^2 + \frac{c\left(p+1\right)}{p+3}v^{\frac{p+3}{p+1}} + Av + B.
    \label{final TW ODE}
\end{equation}
where $A, B$ are two constants of integration.
We then compute the associated traveling solitary solutions on a zero background. 
To this end, we require that $\lim_{Z\to\pm\infty} = 0$ so that $A = B = 0$ and hence Eq.~\eqref{final TW ODE} now becomes
\begin{equation}
    \frac{\sqrt{p}\epsilon^2}{24\left(p+1\right)}\left(v_Z\right)^2 = -\frac{\sqrt{p}}{p+1}v^2 + \frac{c\left(p+1\right)}{p+3}v^{\frac{p+3}{p+1}}.
    \label{traveling soliton ODE}
\end{equation}
We notice that the co-traveling frame ODE \eqref{traveling soliton ODE} does not always admit an analytical solution, 
However, for some specific cases of $p$ including $p = 3, 2, \frac{3}2$, the analytical solutions exist and are given as follows,
\bse
\begin{align}
    &R_{p = 3}\left(Z\right) = -\frac{a_3}{2a_2}\left[\sin\left(\frac{\sqrt{-a_2}}{\sqrt{4a_1}}\left(Z - Z_0\right)\right) + 1\right],
    \label{e:TW solution for p = 3}\\
    &R_{p = 2}\left(Z\right) = \left(\frac{a_3}{2a_2}\right)^2\left[\sin\left(\frac{\sqrt{-a_2}}{\sqrt{\widetilde{a}_1}}\left(Z-Z_0\right)\right) + 1\right]^2,\label{e:TW solution for p = 2}\\
    &R_{p = \frac{3}2}\left(Z\right) = \left(\frac{a_3}{2a_2}\right)^{4}\left[\sin\left(\frac{\sqrt{-a_2}}{2\sqrt{\widetilde{a}_{1}}}\left(Z - Z_0\right)\right) + 1\right]^{4},
    \label{e:TW solution for p = 1.5}  
\end{align}
\ese
where $a_1 = {\sqrt{p}\epsilon^2}/[{24\left(p+1\right)}]$, $ \widetilde{a}_1 = (p+1)^2a_1$, $a_2 = -\sqrt{p}/(p + 1)$, and $a_3 = c(p+1)/(p+3)$. 
It is important to note that these are not solitary but rather periodic 
wave solutions, as constructed.
Hence, when referring to solitary waves here, we mean them in a similar way 
to earlier works such as~\cite{Nester2001,Ahnert_2009}, where a single interval
of positive values between two zeros
of the periodic solution is ``glued'' with zeros on both sides to constitute
an approximation to the relevant solitary wave.

\paragraph{Solitary waves in the regularized model.}

We now turn to the derivation of the solitary waves of the regularized continuum model \eqref{e:BBM-like PDE}. 
Similarly to before, we substitute the ansatz \eqref{Traveling-wave ansatz} into~\eqref{e:BBM-like PDE} and integrate the resulting ODE twice to obtain that
\begin{equation}\label{e:Traveling ODE}
    \frac{\epsilon^2c}{24}\left(R'\right)^2 = cR^2 - \frac{8\sqrt{p}}{\left(p+1\right)\left(p+3\right)}R^{\frac{p+3}2} + 2DR + E,
\end{equation}
where $D, E$ are two constants of integration. 

We then introduce the change of dependent variable $R = \phi^2$ and insert it into the Eq.~\eqref{e:Traveling ODE} to obtain that,
\begin{equation}\label{e:BBM traveling ode}
    \frac{\epsilon^2c}{6}\phi^2\left(\phi'\right)^2 = c\phi^{4} - \frac{8\sqrt{p}}{\left(p+1\right)\left(p+3\right)}\phi^{p+3}+2a\phi^2 + b.
\end{equation}
To compute the solitary wave on the zero background, we set $D = E = 0$, and then integrating Eq.~\eqref{e:BBM traveling ode} yields
\begin{equation}
    \phi\left(Z\right) = \left(\frac{c\left(p+1\right)\left(p+3\right)}{8\sqrt{p}}\right)^{\frac{1}{p-1}}\text{sech}^{\frac2{p-1}}\left(\frac{\sqrt{6}\left(p-1\right)}{2\epsilon}\left(Z-Z_0\right)\right),
\end{equation}
where $Z_0$ is a constant of integration. Finally, we recall that since $R = \phi^2$, the traveling solitary wave solutions read,
\begin{equation}\label{e:Traveling wave solutions}
    R\left(Z\right) = \left(\frac{c\left(p+1\right)\left(p+3\right)}{8\sqrt{p}}\right)^{\frac2{p-1}}\text{sech}^{\frac{4}{p-1}}\left(\frac{\sqrt{6}\left(p-1\right)}{2\epsilon}\left(Z-Z_0\right)\right).
\end{equation}

\begin{figure}[t        !]
    \centering
    \includegraphics[width=0.8\linewidth]{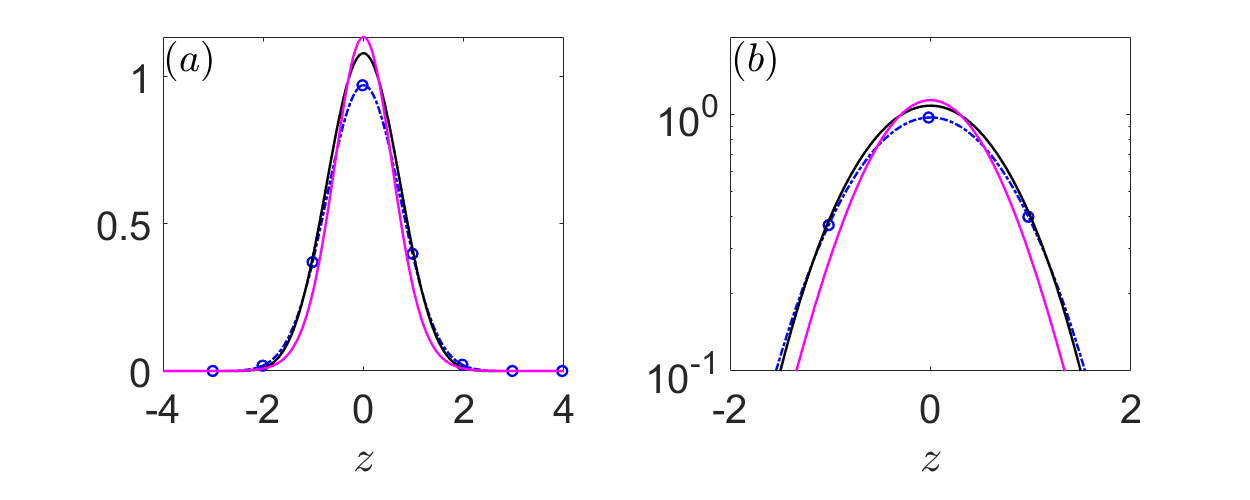}
    \caption{Comparison of the solitary waves for the case of $p = 3/2$. \textbf{(a)} The spatial profiles of the solitary waves and \textbf{(b)} The semi-log plot of all solitary waves. Notice that the blue dashed-dotted line depicts the exact solitary wave of the granular discrete model \eqref{Granular Crystals}, while the solid black and magenta lines showcase the solitary waves of the two continuum models of \eqref{e:KdV-like PDE} and \eqref{e:BBM-like PDE}, respectively. In addition, also note that the solitary wave of the non-regularized model \eqref{e:KdV-like PDE} is only plotted for one period associated with solitary-wave solution in Eq.~\eqref{e:TW solution for p = 1.5}, as per our relevant
    commentary in the text.}
    \label{fig:Solitary waves comparison}
\end{figure}

\paragraph{Comparison of the solitary waves.}

Finally, we compare the solitary waves of the two continuum models with the exact one from the original discrete model \eqref{Granular Crystals}. 
For brevity, we limit ourselves to the value $p=3/2$. 
We notice that if we write the solitary wave solution~\eqref{e:TW solution for p = 1.5} of the non-regularized model in terms of the original granular strain variable, it becomes (ignoring the arbitrary shift parameter $Z_0$)
\begin{align}
    r_{p = \frac{3}2}\left(n,t\right) &= \left(\frac{a_3}{2a_2}\right)^{4}\left[\sin\left(\frac{\sqrt{-6a_2}}{p^{\frac{1}{4}}\left(p+1\right)^{\frac{1}2}}\left(n-ct\right)\right) + 1\right]^{4}.
\label{e:TW solution for p = 1.5 in strain var}
\end{align}
For the regularized continuum model, rewriting the solitary wave solution~\eqref{e:Traveling wave solutions} in terms of the lattice variable $z$ yields
\begin{equation}\label{e:Soliton solution for the regularized model in z}
    r(n,t) = \left(\frac{c(p+1)(p+3)}{8\sqrt{p}}\right)^{\frac2{p-1}}\text{sech}^{\frac{4}{p-1}}\left(\frac{\sqrt{6}(p-1)}2(n-ct)\right).
\end{equation}
Notice from the expressions in~\eqref{e:TW solution for p = 1.5 in strain var} and \eqref{e:Soliton solution for the regularized model in z} that the solitary wave approximations for both continuum models are all independent of the smallness parameter $\epsilon$.

Next, we wish to compare these solitary waves with the one from the discrete granular lattice \eqref{Granular Crystals}. To this end, by applying the co-traveling transform $r_n(t) = r\left(n-ct\right) = r\left(z\right)$, we end up with the following advance-delay equation,
\begin{equation}\label{e:TW ODE for the discrete system}
    c^2r_{zz} = r^{p}\left(z-1\right) - 2r^{p}\left(z\right) + r^{p}\left(z + 1\right).
\end{equation}
The equation \eqref{e:TW ODE for the discrete system} can be numerically solved by the iterative algorithm proposed in \cite{HOCHSTRASSER}.
Figure~\ref{fig:Solitary waves comparison} shows a detailed comparison of the solitary waves of the two continuum models (\eqref{e:KdV-like PDE} and \eqref{e:BBM-like PDE}) and that of the discrete granular lattice \eqref{Granular Crystals} is shown for the case of $p = 3/2$. 
Note that the solitary-wave solution of the non-regularized model~\eqref{e:TW solution for p = 1.5 in strain var} is plotted over only one period, as discussed above.

\subsection{Periodic traveling wave solutions}
\label{s:periodic solutions}

In this section, we discuss briefly the periodic traveling wave solutions of the two continuum models. The periodic traveling wave solutions play an important role in the analysis of the dispersive shock waves, since the latter simply represents the modulated version of the former. However, the periodic solutions for both continuum models do not usually admit analytical expressions. Fortunately this obstacle does not impede our analysis of the DSW. 

\begin{figure}[t!]
    \centering
    \includegraphics[width=0.4\linewidth]{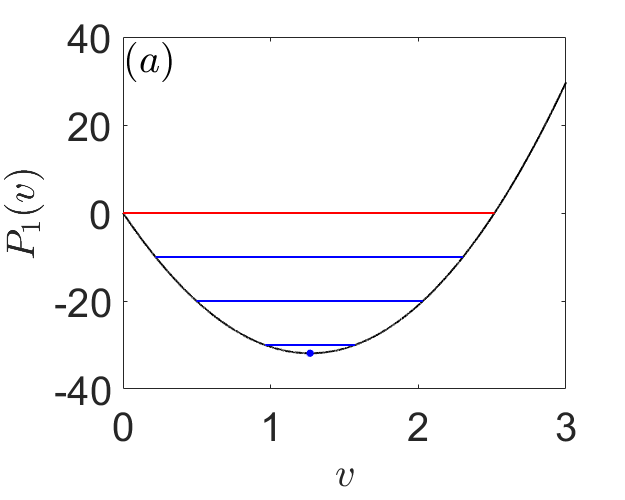}
    \includegraphics[width=0.4\linewidth]{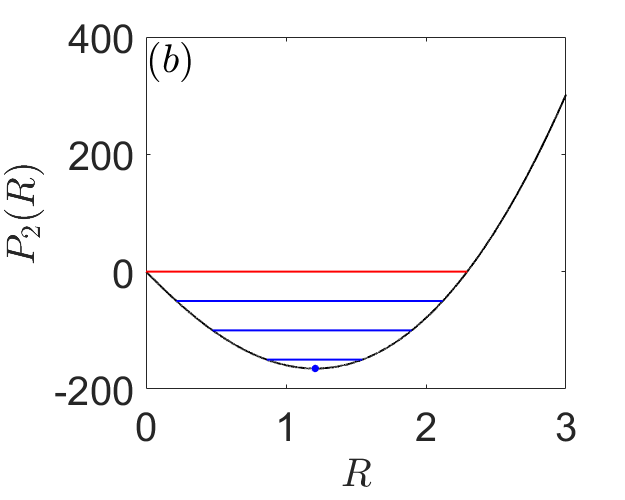}
    \caption{Potential curves. Panels~\textbf{(a)} and~\textbf{(b)} panels the potential curve \eqref{e:potential curve of the Non-regularized model} of the non-regularized model \eqref{e:KdV-like PDE} and that \eqref{e:Potential curve of the regularized model} of the regularized model \eqref{e:BBM-like PDE}. Notice that the red horizontal lines correspond to the solitary-wave solutions of the models, while the three blue horizontal lines in each panel below the red one represent the associated periodic solutions of the models.}
    \label{fig:Potential curves plot}
\end{figure}

\paragraph{Periodic solutions of the non-regularized model.} 
For the non-regularized continuum model \eqref{e:KdV-like PDE}, unfortunately, the periodic solutions are not analytically obtainable. However, we can demonstrate their existence based upon some simple phase plane analysis. In particular, we can visualize the potential curve of the non-regularized co-traveling frame ODE in~\eqref{final TW ODE}, which reads as follows:
\begin{equation}
\label{e:potential curve of the Non-regularized model}
    P_1\left(v\right) = \frac{\sqrt{p}}{p+1}v^2 - \frac{c\left(p+1\right)}{p+3}v^{\frac{p+3}{p+1}} - Av - B.
\end{equation}
The left panel in Fig. \ref{fig:Potential curves plot} shows the potential curve of \eqref{e:potential curve of the Non-regularized model} for the case of $p = 3/2$. 
We first notice that all real solutions $v$ should satisfy that $P_1(v) \leq 0 $, and that each of the three blue characteristic curves represents a distinct periodic solution to the model \eqref{e:KdV-like PDE}. 

\paragraph{Periodic solutions of the regularized model: General case.}
Similarly, for the regularized model \eqref{e:BBM-like PDE}, the associated potential curve, which is visualized in the right panel of Fig.~\ref{fig:Potential curves plot} for the case of $p = 3/2$, reads
\begin{equation}\label{e:Potential curve of the regularized model}
    P_2(R) = \frac{24}{\epsilon^2c}\bigg(-cR^2+\frac{8\sqrt{p}}{(p+1)(p+3)}-2DR-E\bigg).
\end{equation}
The three blue horizontal lines in the right panel of Fig.~\ref{fig:Potential curves plot} also demonstrate the existence of the periodic solutions in the regularized continuum model \eqref{e:BBM-like PDE}. 
However, unlike the non-regularized model \eqref{e:KdV-like PDE}, the regularized model \eqref{e:BBM-like PDE} admits analytical periodic solutions for two particular cases: $p = 3$ and $p = 5$, as we demonstrate next.

\paragraph{Periodic solutions of the regularized model:  $p = 3$.}
When $p = 3$, the co-traveling frame ODE becomes
\begin{equation}
    \frac{\epsilon^2c}{24}\left(R'\right)^2 = cR^2 - \frac{8\sqrt{p}}{\left(p+1\right)\left(p+3\right)}R^{3} + 2aR + b.
\end{equation}
so that
\begin{equation}\label{e:periodic ODE}
    \left(R'\right)^2 = \frac{24a_1}{\epsilon^2c}\left(-R^{3} + \frac{c}{a_1}R^2 + \frac{2a}{a_1}R + \frac{b}{a_1}\right)
    = \frac{24a_1}{\epsilon^2c}\left(R_1-R\right)\left(R_2-R\right)\left(R_3-R\right),
\end{equation}
where $a_1 = {8\sqrt{p}}/[{\left(p+1\right)\left(p+3\right)}]$, and $R_1 \leq R_2 \leq R_3$ are the three roots of the polynomial $P\left(R\right) = -a_1 R^{3} +c\,R^2 + 2a\,R + b$.
A direct integration of~\eqref{e:periodic ODE} yields
\begin{equation}\label{e:periodic solution for p = 3}
    R\left(Z\right) = R_2 + \left(R_3 - R_2\right)\text{cn}^2\left(\frac{\sqrt{6\left(R_3-R_1\right)a_1}}{\sqrt{c}\epsilon}\left(Z-Z_0\right), m\right),
\end{equation}
where $m = (R_3 - R_2)/(R_3 - R_1)$ is the elliptic parameter, $\text{cn}$ denotes the Jacobi elliptic cosine, and $Z_0$ is a constant of integration.
Note that, in the soliton limit, $m \to 1$, or equivalently $R_2 \to R_1$, we can further compute the soliton amplitude by noticing that
\begin{equation}\label{e:expansion of the product}
    R_1 + R_2 + R_3 = {c}/{a_1}.
\end{equation}
Therefore, in the soliton limit we have $R_3 = {c/}{a_1} - 2R_1 = {c/}{a_1} - 2r^+$, where $r^+$ denotes the background value. Moreover, the soliton amplitude is 
$a^+ = ({s^+}/{a_1}) - 3r^+$.

\paragraph{Periodic solutions of the regularized model: $p = 5$.}
On the other hand, when $p = 5$, the co-traveling frame ODE reads
\begin{equation}
    \begin{aligned}
    \frac{\epsilon^2c}{24}\left(R'\right)^2 &= cR^2 - a_1 R^{4} + 2aR + b.
    \end{aligned}
\end{equation}
We then have
\begin{equation}
\label{e:co-traveling ODE for p = 5}
        \left(R'\right)^2 = \frac{24}{\epsilon^2c}\left(cR^2 - a_1R^{4} + 2aR + b\right)
        = -\frac{24a_1}{\epsilon^2c}\left(R-R_1\right)\left(R-R_2\right)\left(R-R_3\right)\left(R-R_4\right),
\end{equation}
where $R_1 \le R_2 \le R_3 \le R_4$ denote the four roots of the polynomial $P\left(R\right) = a_1 R^{4} - {c}\,R^2 - {2a}\,R - {b}$.

Let $\mu = -{24a_1}/({\epsilon^2c})$.
If $\mu > 0$, a direct integration of the co-traveling frame ODE \eqref{e:co-traveling ODE for p = 5} yields
\begin{equation}
\label{e:periodic wave ass. positive dispersion}
    R = R_2 + \frac{\left(R_3 - R_2\right)\text{cn}^2\left(\zeta, m\right)}{1 - \frac{R_3 - R_2}{R_4 - R_2}\text{sn}^2\left(\zeta, m\right)},
\end{equation}
where 
\begin{equation}
\label{e:def of phase and elliptic modulus}
        \zeta = \frac{\sqrt{\left|\mu\right|\left(R_3 - R_1\right)\left(R_4 - R_2\right)}}2Z,\qquad
        m = \frac{\left(R_3 - R_2\right)\left(R_4 - R_1\right)}{\left(R_4 - R_2\right)\left(R_3 - R_1\right)}.
\end{equation}
Similarly to the previous case, we can then make theoretical predictions about the soliton amplitude based on the soliton limit of~\eqref{e:periodic wave ass. positive dispersion}. 
In this case, however, the soliton limit $m \to 1$ can be reached via two ways: either $R_2 \to R_1$ or $R_3 \to R_4$. For the former case, the periodic solution~\eqref{e:periodic wave ass. positive dispersion} reduces to
\begin{equation}
    R = R_1 + \frac{R_3 - R_1}{\text{cosh}^2\zeta - \frac{R_3 - R_1}{R_4 - R_1}\text{sinh}^2\zeta},
\end{equation}
and the soliton amplitude thereof is $a^+ = R_3 - R_1$.
Conversely, in the latter case of $R_3 \to R_4$, the periodic solution now reduces to
\begin{equation}
    R = R_4 - \frac{R_4 - R_2}{\text{cosh}^2\zeta - \frac{R_4 - R_2}{R_4 - R_1}\text{sinh}^2\zeta},
\end{equation}
from which we infer that the soliton amplitude is now $a^+ = R_4 - R_2$.

Now, we investigate the case when $\mu < 0$. In this situation, the periodic wave oscillates either in the interval $R_1 \le R \le R_2$ or in $R_3 \le R \le R_4$. We assume the latter case but the periodic wave solution for the former case can be derived analogously. We integrate the co-traveling frame ODE \eqref{e:co-traveling ODE for p = 5} to obtain that
\begin{equation}\label{e:periodic wave solution ass. negative dispersion}
    R = R_3 + \frac{\left(R_4 - R_3\right)\text{cn}^2\left(\zeta, m\right)}{1 + \frac{R_4 - R_3}{R_3 - R_1}\text{sn}^2\left(\zeta, m\right)},
\end{equation}
where $\zeta$ is still given by~\eqref{e:def of phase and elliptic modulus}, 
but where the elliptic modulus $m$ is given as follows
\begin{equation}\label{e:elliptic modulus m2}
    m = \frac{\left(R_4 - R_3\right)\left(R_2 - R_1\right)}{\left(R_4 - R_2\right)\left(R_3 - R_1\right)}.
\end{equation}
In the soliton limit, $m \to 1$, we have that $R_3 \to R_2$, and the periodic solution~\eqref{e:periodic wave solution ass. negative dispersion} reduces to
\begin{equation}
    R = R_2 + \frac{R_4 - R_2}{\text{cosh}^2\zeta + \frac{R_4 - R_2}{R_2 - R_1}\text{sinh}^2\zeta},
\end{equation}
so that the soliton amplitude reads
$a^+ = R_4 - R_2$.
To obtain an explicit formula for the soliton amplitude $a^+$ from this expression, 
we notice that by expanding the product of $\left(R-R_1\right)\left(R-R_2\right)\left(R-R_3\right)\left(R-R_4\right)$ and equating the coefficients with that of the polynomial $P(R)/a_1$, we end up with the following constraints:
\begin{equation}
        R_1 + R_2 + R_3 + R_4 = 0,\qquad
        R_1R_2 + R_1R_3 + R_2R_3 + R_1R_4 + R_2R_4 + R_3R_4 = - {c}/{a_1}.
\end{equation}
Since $R_3=R_2$ in the soliton limit, we then solve for $R_1$ and $R_4$.  
A direct substitution of them into the expression for $a_+$ then yields
\begin{equation}
    a^+ =  - 2r^+ + \sqrt{(c/a_1) - 2(r^+)^2},
\end{equation}
where $r^+ = R_2$ is the background on top of which the soliton rises.

\section{Conservation laws and Whitham modulation equations}
\label{s:Whitham equations}

In this section, we will derive the so-called Whitham modulation equations for both the non-regularized and the regularized continuum models derived in Section~\ref{s:models}. 
Recall that the Whitham modulation equations are a system of first-order, hydrodynamic-type PDEs which govern the spatio-temporal evolution of the parameters of the periodic traveling waves of the underlying PDE. 
As we will show in the later sections, the modulation systems also capture the evolution features of the dispersive shocks, and they hence play a fundamental role in the theoretical analysis of the profile of DSWs.

\subsection{Conservation laws for the continuum models}
\label{s:conservation laws}

As a prerequisite for the derivation of the Whitham modulation equations by averaged conservation laws, we need to find conservation laws for the two continuum models~\eqref{e:KdV-like PDE} and \eqref{e:BBM-like PDE}. Notice that the periodic traveling waves of both the non-regularized and regularized continuum models contain three parameters. This suggests that we need two conservation laws per continuum model to construct closed Whitham modulation systems (since one of the modulation equations is simply the so-called conservation of waves). 

\paragraph{Conservation laws for the non-regularized model.}
Instead of working on the PDE at the level of the strain $r$, 
it is convenient to employ the change of dependent variable $v(X,T) = r^{(p+1)/2}$, 
whereby the PDE \eqref{e:KdV-like PDE} is mapped onto
\begin{equation}
    \left(v^{\frac2{p+1}}\right)_{T} + \frac{2\sqrt{p}}{p+1}v_{X} + \frac{\sqrt{p}\epsilon^2}{12\left(p+1\right)}v_{XXX}= 0.
    \label{PDE for v}
\end{equation}
We then notice that a straightforward rearrangement of this PDE yields the following first conservation law:
\begin{equation}
\label{First conservation law}
    \left(v^{\frac2{p+1}}\right)_T + \left(\frac{2\sqrt{p}}{p+1}v + \frac{\sqrt{p}\epsilon^2}{12\left(p+1\right)}v_{XX}\right)_X= 0.
\end{equation}
For the second conservation law, by multiplying the PDE \eqref{PDE for v} by $v$, we obtain
\begin{equation}
\label{second conservation law}
    \frac2{p+3}\left(v^{\frac{p+3}{p+1}}\right)_T + \left(\frac{\sqrt{p}}{p+1}v^2 + \frac{\sqrt{p}\epsilon^2}{24\left(p+1\right)}\left(2vv_{XX} - v_{X}^2\right)\right)_X = 0,
\end{equation}
which gives the second conservation law.
Finally, rewriting the two conservation laws in terms of the field variable $r$ yields
\bse
\begin{align}
    r_T + \left(\frac{2\sqrt{p}}{p+1}r^{\frac{p+1}2} + \frac{\sqrt{p}\epsilon^2}{12\left(p+1\right)}\left(r^{\frac{p+1}2}\right)_{XX}\right)_{X} &= 0,
    \label{conservation law 1 in terms of r}\\
    \frac2{p+3}\left(r^{\frac{p+3}2}\right)_T + \left(\frac{\sqrt{p}}{p+1}r^{p+1} + \frac{\sqrt{p}\epsilon^2}{24\left(p+1\right)}\left(2r^{\frac{p+1}2}\left(r^{\frac{p+1}2}\right)_{XX} - \left[\left(r^{\frac{p+1}2}\right)_{X}\right]^2\right)\right)_X &= 0.
    \label{conservation law 2 in terms of r}
\end{align}
\ese
Note that \eqref{First conservation law} corresponds to the conservation of mass,
and \eqref{conservation law 2 in terms of r} to the conservation of momentum.

\paragraph{Conservation laws for the regularized model.}
Proceeding in a similar way for the regularized continuum model \eqref{e:BBM-like PDE}, one obtains the following two conservation laws:
\bse
\begin{align}
    \left(r - \frac{\epsilon^2}{24}r_{XX}\right)_{T} + \frac{2\sqrt{p}}{p + 1}\left(r^{\frac{p+1}2}\right)_{X} &= 0,\label{e:first conservation law}\\
    \left(\frac{1}2r^2 + \frac{\epsilon^2}{48}\left(r_X\right)^2\right)_{T} + \left(\frac{2\sqrt{p}}{p+3}r^{\frac{p+3}2} - \frac{\epsilon^2}{24}rr_{XT}\right)_{X} &= 0,\label{e:second conservation law}
\end{align}
\ese
Again, \eqref{e:first conservation law} corresponds to the conservation of mass, while \eqref{e:second conservation law} to the conservation of momentum.

\subsection{Modulation system for the non-regularized model}
\label{s:Whitham_nonregularized}

Next, we apply the method of averaging the conservation laws in order to derive the modulation equations for the non-regularized model~\eqref{e:KdV-like PDE}.
To this end, let $\phi\left(\theta\right)$ denote a periodic traveling wave solution with a fixed period $2\pi$, where 
\begin{equation}
\theta = \left(KX - \Omega T\right)/\epsilon
\label{e:thetadef}
\end{equation}
denotes a fast phase. 
We seek a slowly modulated wave in the following form:
\begin{equation}
\label{e:slowly modulated wave}
    r\left(X,T\right) = \phi\left(\theta\right) + \epsilon r_1\left(\theta\right) + \cdots, \hspace{4mm} 0 < \epsilon \ll 1,
\end{equation}
where we recall that $\epsilon$ is the formal smallness parameter we used before,
and where the fast phase variable $\theta$ is now defined (up to an inessential translation constant) by the relations
\begin{equation}
\theta_X = K/\epsilon, \hspace{5mm} \theta_T = -\Omega/\epsilon\,,
\label{e:KOmegadef}
\end{equation}
with the local wavenumber $K$ and the local frequency $\Omega$ now slowly varying functions of $X$ and~$T$.
Then, the compatibility condition $\theta_{XT} = \theta_{TX}$ immediately yields the 
so-called conservation of waves condition:
\begin{equation}
\label{e:conservation of waves in terms of X and T}
    K_{T} + \Omega_{X} = 0,
\end{equation}
which is the first modulation equation.
Next, to apply the method of averaging the conservation laws, we introduce the averaging operator
\begin{equation}
    \overline{F\left(\phi\right)} = \frac{1}{2\pi}\int_0^{2\pi} F\left(\phi\left(\theta\right)\right)d\theta.
    \label{Averaging operation}
\end{equation}
We then insert our slowly modulated wave \eqref{e:slowly modulated wave} into the first conservation law \eqref{conservation law 1 in terms of r} and apply the averaging operation \eqref{Averaging operation} to obtain
\begin{equation}
\label{e:modulation eq 2}
    \big(\overline\phi\big)_T + \frac{2\sqrt{p}}{p+1}\left[\left(\overline{\phi^{\frac{p+1}2}}\right)\right]_{X} = \mathcal{O}\left(\epsilon\right),
\end{equation}
where we used the fact that the averaging operation and the partial differentiation with respect to $X$ and $T$ commute at the leading order due to the fixed $2\pi$ constant period of our periodic wave solutions.
Similarly, substituting the slowly periodic wave \eqref{e:slowly modulated wave} into the second conservation law \eqref{conservation law 2 in terms of r}, we obtain
\begin{equation}\label{e:modulation eq 3}
    \frac2{p+3}\left(\overline{\phi^{\frac{p+3}2}}\right)_T + 
    \left(\frac{\sqrt{p}}{p+1}\overline{\phi^{p+1}} - \frac{\sqrt{p}K^2}{8\left(p+1\right)}\overline{\left[\left(\phi^{\frac{p+1}2}\right)_{\theta}\right]^2}\right)_X = \mathcal{O}\left(\epsilon\right).
\end{equation}
Dropping higher order terms in $\epsilon$, equations 
\eqref{e:conservation of waves in terms of X and T}, \eqref{e:modulation eq 2} and together with \eqref{e:modulation eq 3}  form the following closed Whitham modulation system
for the periodic traveling solutions of the non-regularized model \eqref{e:KdV-like PDE}:
\begin{subequations}
\begin{align}
    K_T + \Omega_X &= 0,
    \label{e:final system of me1}\\
    \overline\phi_T + \frac{2\sqrt{p}}{p+1}\left(\overline{\phi^{\frac{p+1}2}}\right)_{X} &= 0,
    \label{e:final system of me2}\\
    \frac2{p+3}\left(\overline{\phi^{\frac{p+3}2}}\right)_T + 
    \left(\frac{\sqrt{p}}{p+1}\overline{\phi^{p+1}} - \frac{\sqrt{p}K^2}{8\left(p+1\right)}\overline{\left[\left(\phi^{\frac{p+1}2}\right)_{\theta}\right]^2}\right)_X &= 0.
    \label{e:final system of me3}
\end{align}
\end{subequations}

\subsection{Modulation system for the regularized model}

\paragraph{Reparametrization of the periodic traveling wave solutions.}

As in section~\ref{s:Whitham_nonregularized}, a prerequisite for the derivation of the modulation equations is to have period traveling wave solutions with fixed period. In the previous
section, we derived an abstract form of the modulation equations, since
no analytical formula for the periodic waves was available. 
In that derivation, the abstract periodic wave was assumed to be
of fixed period. One can proceed in a similar way for the regularized model.
However, since we have an exact formula for the periodic wave (e.g., in the
case of $p=3$),
we derive the modulation equations more explicitly. The first step
is to 
reparametrize the periodic traveling wave solutions to have a fixed period, in preparation for the derivation of the modulation equations, dealing for example with the solutions in~\eqref{e:periodic solution for p = 3}
for $p=3$. 
To this end, we introduce the same fast phase variable $\theta$ as in~\eqref{e:thetadef},
with $\Omega = cK$.
In terms of $\theta$, the periodic traveling wave solution reads
\begin{equation}
\label{e:periodic wave in fast variable}
    R\left(\theta\right) = R_2 + \left(R_3 - R_2\right)\text{cn}^2\left(\frac{\sqrt{6\left(R_3 - R_1\right)a_1}}{\sqrt{c}K}\theta, m\right),
\end{equation}
where $m = (R_3 - R_2)/(R_3 - R_1)$ as before, 
and where we have ignored an arbitrary translation constant~$\theta_0$.
Then we observe that, since the periodic wave oscillates between the roots $R_2$ and~$R_3$,
we have
\begin{equation}
\label{e:change of period calculation}
    2\pi = \int_{0}^{2\pi}d\theta = 2\int_{R_2}^{R_3}\frac{dR}{R_{\theta}} = \frac{4\sqrt{K^2c}K_m}{\sqrt{24a_1\left(R_3-R_1\right)}},
\end{equation}
where $K_m = K(m)$ (not to be confused with the wavenumber $K$) denotes the complete elliptic integral of the first kind.
Therefore, from~\eqref{e:change of period calculation} we deduce that
$\sqrt{6\left(R_3-R_1\right)a_1}/(\sqrt{c} K) = K_m/\pi$,
and substituting this relation into the original periodic wave solution \eqref{e:periodic wave in fast variable} we obtain 
\begin{equation}
\label{e:periodic wave with 2pi fixed period}
    R\left(\theta\right) = R_2 + \left(R_3 - R_2\right)\text{cn}^2\left(\frac{K_m}{\pi}\theta, m\right),
\end{equation}
confirming that the periodic traveling wave solution~\eqref{e:periodic wave with 2pi fixed period} has a fixed period of $2\pi$.

Next, we observe that the parameters $R_1, R_2, R_3$ are in one-to-one correspondence with $c, K, m$. 
Indeed, this can be readily seen based on the following set of relations:
\begin{equation}
\label{e:relations between roots and c,K,m}
        R_3 - R_1 = \frac{cK^2K_m^2}{6a_1\pi^2},\qquad
        R_1 + R_2 + R_3 = \frac{c}{a_1},\qquad
        \frac{R_3 - R_2}{R_3 - R_1} = m.
\end{equation}
If desired, one can solve the system \eqref{e:relations between roots and c,K,m} for $R_1, R_2, R_3$ in terms of $c,K,m$, to obtain
\begin{subequations}
\label{e:solutions for R1 R2 R3}
\begin{gather}
      R_1 = \frac{6a_1\pi^2c - 2a_1cK^2K_m^2+a_1cK^2K_m^2m}{18a_1^2\pi^2},\\
      R_2 = \frac{6a_1\pi^2c+a_1cK^2K_m^2-2a_1cK^2K_m^2m}{18a_1^2\pi^2},\\
      R_3 = \frac{6a_1\pi^2c + a_1cK^2K_m^2+a_1cK^2K_m^2m}{18a_1^2\pi^2}.
\end{gather}
\end{subequations}
Finally, one can substitute~\eqref{e:solutions for R1 R2 R3} into~\eqref{e:periodic wave with 2pi fixed period} to obtain
a $2\pi$ fixed-period three-parameter family of periodic traveling wave solutions in terms of the parameters of $c, K, m$.
The explicit expression is omitted for brevity.

\paragraph{Derivation of the modulation equations.}

We apply the method of averaging the conservation laws to derive the Whitham modulation equations. Notice first that the periodic wave is always parametrized by three parameters of $a, b, c$, so this suggests we need three modulation equations to describe the slowly modulational behaviors of the parameters. To this end, we seek a slowly modulated periodic wave in the same form as~\eqref{e:slowly modulated wave},
where the fast phase $\theta$ is still defined by the conditions~\eqref{e:KOmegadef},
so that the local wavenumber $K$ and local frequency $\Omega$ still satisfy the 
conservation of waves relation~\eqref{e:conservation of waves  in terms of X and T}.
We then average the two conservation laws in section \ref{s:conservation laws} over a period, to obtain 
\bse
\begin{align}
    \overline\phi_{T} + \frac{2\sqrt{p}}{p + 1}\left(\overline{\phi^{\frac{p+1}2}}\right)_{X} &= \mathcal{O}\left(\epsilon\right),\label{e:second me in higher-order form}\\
    \left(\frac{1}2\overline{\phi^2} + \frac{K^2}{48}\overline{\phi_\theta^2}\right)_{T} + \left(\frac{2\sqrt{p}}{p+3}\overline{\phi^{\frac{p+3}2}} - \frac{\Omega K}{24}\overline{\phi_\theta^2}\right)_X &= \mathcal{O}\left(\epsilon\right).\label{e:third me in higher-order form}
\end{align}
\ese
Then, by dropping the higher order terms in $\epsilon$, and including the conservation of waves equation, we finally arrive at
\bse
\begin{align}
K_T + \Omega_X &= 0,\label{e:first me} \\
    \overline\phi_{T} + \frac{2\sqrt{p}}{p + 1}\left(\overline{\phi^{\frac{p+1}2}}\right)_{X} &= 0,\label{e:second me}\\
    \left(\frac{1}2\overline{\phi^2} + \frac{K^2}{48}\overline{\phi_\theta^2}\right)_{T} + \left(\frac{2\sqrt{p}}{p+3}\overline{\phi^{\frac{p+3}2}} - \frac{\Omega K}{24}\overline{\phi_\theta^2}\right)_X &= 0.\label{e:third me}
\end{align}
\ese

\subsection{Harmonic and solitonic reductions of the modulation systems}

In this section we consider the reduction of the modulation equations for both the non-regularized and regularized PDE models in both the harmonic and the solitonic limit,
which will be useful when characterizing the DSWs of these systems.
We first notice that at both the harmonic and solitonic limits, one has the following relations:
\begin{equation}
\label{e:two important facts in the edge reduction}
        \overline{\phi^{n}} = \big(\overline\phi\big)^{n},\qquad
        \overline{\phi_\theta^2} \to 0. 
\end{equation}
As a consequence, the modulation equations in the harmonic limit simply reduce to the following two equations:
\begin{equation}
\label{e:reduced modulation system}
    \begin{aligned}
        K_T + \left(\Omega_0\right)_X &= 0,\\
        \big(\overline\phi\big)_T + \sqrt{p}\big(\overline\phi\big)^{\frac{p-1}2}\big(\overline\phi\big)_X &= 0,
    \end{aligned}
\end{equation}
where $\Omega_0$ denotes the linearized dispersion relation, namely~\eqref{e:final LDR} in the case of the non-regularized modulation equations and~\eqref{eq:LDR for BBM} for the regularized ones.
The system \eqref{e:reduced modulation system} can be written in the following matrix form:
\begin{equation}
\label{e:matrix form of the reduced system}
    \begin{bmatrix}
        K \\ \overline\phi
    \end{bmatrix}_{T}
    + 
    \begin{bmatrix}
        {\partial\Omega_0}/{\partial K} & {\partial\Omega_0}/{\partial \overline\phi} \\
        0 & \sqrt{p}\big(\overline\phi\big)^{\frac{p-1}2}
    \end{bmatrix}
    \begin{bmatrix}
        K \\
        \overline\phi
    \end{bmatrix}_X 
    = 0.
\end{equation}
The coefficient matrix in~\eqref{e:matrix form of the reduced system} 
has the following two left eigenvectors:
\begin{equation}
\label{e:two left eigenvectors}
        v_{1} = \left[0,1\right],\qquad
        v_2 = \left[\frac{\partial \Omega_0}{\partial K} - \sqrt{p}\big(\overline\phi\big)^{\frac{p-1}2}, \frac{\partial\Omega_0}{\partial\overline\phi}\right],
\end{equation}
associated with the eigenvalues $\lambda_1 = \sqrt{p}\big(\overline\phi\big)^{\frac{p-1}2}$ and $\lambda_2 = {\partial\Omega_0}/{\partial K}$, respectively.
Multiplying the system \eqref{e:matrix form of the reduced system} with the second left eigenvector $v_2$, we then obtain its associated characteristic form,
\begin{equation}\label{e:characteristic form of the system}
    \left(\frac{\partial\Omega_0}{\partial K} - \sqrt{p}\big(\overline\phi\big)^{\frac{p-1}2}\right)\frac{dK}{dT} + \frac{\partial\Omega_0}{\partial \overline\phi}\frac{d\overline\phi}{dT} = 0,
\end{equation}
which further implies 
\begin{equation}
    \frac{dK}{d\overline\phi} = \frac{\partial\Omega_0/\partial\overline\phi}{\sqrt{p}\big(\overline\phi\big)^{\frac{p-1}2} - \partial\Omega_0/\partial K}.
    \label{ODE 1}
\end{equation}
The ODE \eqref{ODE 1} will be useful to characterize the trailing edge of the DSWs, in which case it
will be supplemented with the boundary condition $K\big(\overline\phi^+\big) = 0$, where $\overline\phi^+ = r^+$. 
In other words, the wavenumber is zero at the leading solitary edge of the DSW.

The calculations for the solitonic limit are slightly more complicated, precisely because
the wavenumber is zero at the solitonic edge of the DSW.
In this case, following \cite{El_2005},
we resort to using the so-called ``conjugate dispersion relation'' 
$\widetilde\Omega_s$ defined as follows:
Letting $\widetilde{K}$ be the conjugate wavenumber \cite{El_2005}, the conjugate dispersion
relation $\widetilde\Omega_s$ is defined as
\begin{equation}
    \widetilde\Omega_s\left(\overline{r},\widetilde{K}\right) = -i\Omega_0\left(\overline{r},i\widetilde{K}\right).
\end{equation}
The ODE satisfied by~$\widetilde{K}$ is analogous to \eqref{ODE 1}. 
More specifically, we have that $K$ and $\widetilde{K}$ must satisfy the equations
\begin{subequations}
\label{e:DSW fitting BVP}
\begin{align}
    \frac{dK}{d\overline{r}} &= \frac{ \partial \Omega_0/\partial \overline{r}}{V\big(\overline{r}\big) - \partial \Omega_0/\partial K},\qquad
    K\left(r^+\right) = 0,
\\
    \frac{d\widetilde{K}}{d\overline{r}} &= \frac{\partial \widetilde\Omega_s/\partial \overline{r}}{V\big(\overline{r}\big) - \partial \widetilde\Omega_s/\partial \widetilde{K}}, \qquad
    \widetilde{K}\left(r^-\right) = 0.
\end{align}
\end{subequations}
where 
\begin{equation}\label{e:definition of V}
V\big(\overline{r}\big) = \sqrt{p}\hspace{0.2mm}\overline{r}^{\frac{p-1}2}.
\end{equation}
These relations will be useful to characterize the solitonic edge of the DSWs.

\section{Riemann problems, rarefaction waves and DSW fitting}

\subsection{Riemann problems}
\label{s:Riemann problems}

To study numerically the dispersive shock waves of the two continuum models \eqref{e:KdV-like PDE} and \eqref{e:BBM-like PDE}, we consider the Riemann problems for the two PDEs. 
We use the pseudo-spectral method for the spatial discretization with a fourth-order Runge-Kutta (RK4) scheme in time to integrate numerically both continuum models. 
For the discrete granular lattice simulation, we simply apply an RK4 time stepping. 
Since the pseudo-spectral discretization requires periodic boundary conditions, however, we consider a periodic variant of the Riemann initial data consisting of box-type initial data.
Also, in order to minimize the production of spurious high wavenumbers, we smooth out the transition between the two constant values, resulting in the following box-type initial conditions:
\begin{equation}
\label{e:IC}
   r\left(X,0\right) = r^+ - \frac12(r^+ - r^-)\left[\tanh\left(\delta\left(X-a\right)\right) - \text{tanh}\left(\delta\left(X-b\right)\right)\right],
\end{equation}
where $a$ and $b$ denote to the left and right edge of the initial ``box'', respectively,
and $\delta=50$ determines the sharpness of the transition between $r^-$ and $r^+$.

Note that, in order to compare the dynamics of the two continuum models with those of the granular lattice~\eqref{Granular Crystals}, we must set up the initial data for the lattice appropriately. 
In particular,
the initial condition of $s_n= \dot r_n$ must be consistent with the ICs for the corresponding continuum model. Namely, we notice that, by the chain rule, $\dot r_n(0) = \epsilon r_T(\epsilon n,0)$, and to compare the results with the non-regularized PDE model, the initial condition for $s_n$ must read
\begin{subequations}
\begin{equation}
\label{e:IC for the velocity of the strain of the DDE}
    s_n\left(0\right) = - \frac{2\epsilon\sqrt p}{p+1} \left( \mathcal{F}^{-1}\left[ik\mathcal{F}\left[r\left(\epsilon n\right)^{\frac{p+1}2}\right]\right] + \frac{\epsilon^2}{24}\mathcal{F}^{-1}\left[-ik^{3}\mathcal{F}\left[r\left(\epsilon n\right)^{\frac{p+1}2}\right]\right]\right),
\end{equation}
where $\mathcal{F}$ and $\mathcal{F}^{-1}$ %
denote the Fourier and inverse Fourier transform operator, respectively, and $k$ is the Fourier wavenumber.
Correspondingly, for the regularized continuum model, the initial condition for $s_n$ is
\begin{equation}\label{e:IC for the velocity of the lattice}
    s_n\left(0\right) = - \frac{2\epsilon\sqrt p}{\left(p+1\right)\left(1+\frac{\epsilon^2k^2}{24}\right)} \mathcal{F}^{-1}\left[ik\,\mathcal{F}\left[r\left(\epsilon n\right)^{\frac{p+1}2}\right]\right].
\end{equation}
\end{subequations}
Note that these two ICs coincide at leading order in $\epsilon$.

\subsection{Rarefaction wave}

Because the ICs~\eqref{e:IC} for the two continuum models~\eqref{e:KdV-like PDE} and~\eqref{e:BBM-like PDE} also include an increasing step, they also give rise to a rarefaction wave in addition to a DSW.
In this section, we show how this rarefaction wave can be characterized via the dispersionless limits of both models, both of which are
\begin{equation}
\label{e:dispersionless limit of the model}
    r_T + \sqrt{p}r^{\frac{p-1}2}r_X = 0.
\end{equation}
A rarefaction wave emerges from the evolution of the Cauchy problem for \eqref{e:dispersionless limit of the model} with the following upward Riemann initial data
\begin{equation}\label{e:upward IC}
    r\left(X,0\right) =
    \begin{cases}
        r^-, \hspace{5mm} X\leq 0\\
        r^+, \hspace{5mm} X > 0,
    \end{cases}
\end{equation}
where $r^- < r^+$.
This rarefaction wave can be represented by a self-similar solution of~\eqref{e:dispersionless limit of the model} in the following form:
\begin{equation}
\label{e:self-similar form}
    r\left(X,T\right) = S(\xi), \qquad 
    \xi = X/T.
\end{equation}
Substitution of the self-similar ansatz \eqref{e:self-similar form} into the dispersionless system \eqref{e:dispersionless limit of the model} yields 
\begin{equation}\label{e:self-similar ODE}
    \bigg( -\xi  + \sqrt{p}S^{\frac{p-1}2} \bigg)\,S_\xi = 0.
\end{equation}
We can then solve~\eqref{e:self-similar ODE} together with the initial condition \eqref{e:upward IC} to obtain
\begin{equation}
\label{e:Self-similar solution}
    S\left(X,T\right) =
    \begin{cases}
        r^-, \hspace{17.5mm} X \leq \sqrt{p}(r^+)^{\frac{p-1}2}T,\\
        \left(\frac{X}{\sqrt{p}T}\right)^{\frac2{p-1}}, \hspace{5mm} \sqrt{p}(r^+)^{\frac{p-1}2}T < X \leq \sqrt{p}(r^-)^{\frac{p-1}2}T,\\
        r^+, \hspace{17.5mm} X > \sqrt{p}(r^-)^{\frac{p-1}2}T.
    \end{cases}
\end{equation}

\begin{figure}[t!]
    \centering
    \includegraphics[width=0.8\linewidth]{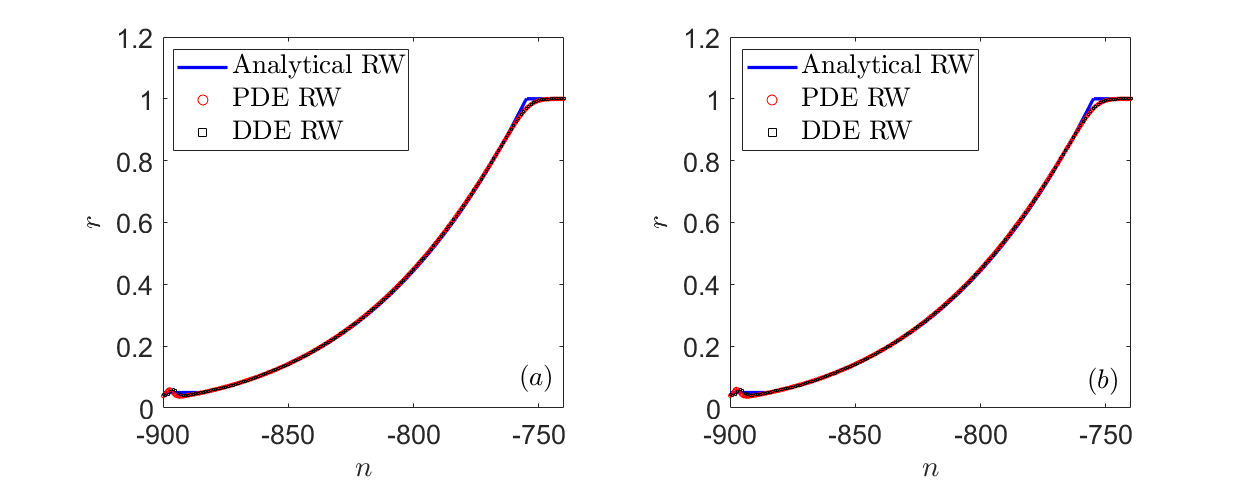}
    \caption{Comparison of the self-similar solution \eqref{e:Self-similar solution} with the numerical rarefaction waves (RWs). Panels \textbf{(a)} and~\textbf{(b)} show respectively the RW of the non-regularized model~\eqref{e:KdV-like PDE} and that of the regularized model~\eqref{e:BBM-like PDE} at $t = 200$ ($T = 20$). 
    The two background values were $r^- = 0.05$ and $r^+ = 1$. 
    The blue solid curve depicts the analytical self-similar solution~\eqref{e:Self-similar solution}, while the red circles and black squares are respectively the numerical RWs of the two continuum PDEs and the associated discrete models.}
    \label{fig:RW comparisons}
\end{figure}

We can now compare the analytical self-similar solution~\eqref{e:Self-similar solution} with the rarefaction waves obtained from numerical simulations of both the non-regularized model \eqref{e:KdV-like PDE}, the regularized model~\eqref{e:BBM-like PDE} and the associated discrete granular chain model~\eqref{Granular Crystals}. 
The results, shown in Fig.~\ref{fig:RW comparisons}, demonstrate excellent agreement between the analytical and numerical rarefaction profiles.
Note also how these profiles are markedly different from the linear ramp that one would obtain for a similar problem in the KdV equation.

\subsection{DSW fitting}
\label{s:DSW fitting}

In this section we apply the so-called DSW fitting method \cite{El_2005} to characterize the leading and trailing edges of the dispersive shock waves.
We first perform DSW fitting on the two continuum models~\eqref{e:KdV-like PDE} and \eqref{e:BBM-like PDE}, and finally on the discrete granular chain \eqref{Granular Crystals}.

\paragraph{Non-regularized model.}

For the non-regularized model~\eqref{e:KdV-like PDE}
we solve~\eqref{e:DSW fitting BVP} with $\Omega_0$ given by~\eqref{e:final LDR}.
This yields
\begin{equation}
    \overline{r} = r^+\left(\frac{1}{1-\frac{1}{24}\epsilon^2K^2}\right)^{\frac{3}{p-1}}, \qquad
    \overline{r} = r^-\left(\frac{1}{1 + \frac{1}{24}\epsilon^2\widetilde{K}^2}\right)^{\frac{3}{p-1}}.
\end{equation}
Then, we notice that the trailing-edge wavenumber $K^{-}$ and the leading-edge conjugate wavenumber $\widetilde{K}^{+}$ are simply obtained as $K^{-} = K(r^{-})$ and $\widetilde{K}^{+} = \widetilde{K}(r^{+})$. 
Namely,
\begin{equation}
    K^- = \frac{\sqrt{24\left(1 - (r^-/r^+)^{\frac{1-p}{3}}\right)}}{\epsilon}, \qquad
    \widetilde{K}^+ = \frac{\sqrt{24\left((r^+/r^-)^{\frac{1-p}{3}}-1\right)}}{\epsilon}
\end{equation}
Furthermore, the trailing and leading-edge velocities, denoted respectively as $s^-$ and $s^+$, are obtained as the phase and group velocities ($\widetilde\Omega_s/\widetilde{K}$ and
$\partial\Omega_0/\partial K$), respectively. Namely, 
\bse
\begin{align}
    s^- &= \frac{\partial \Omega_0}{\partial K}\left(r^-, K^-\right) = \sqrt{p}\left(r^-\right)^{\frac{p-1}2}\left(1 - \frac{1}{8}\epsilon^2\left(K^-\right)^2\right),
    \label{e:non-reg trailing-edge speed}\\
    s^+ &= \frac{\widetilde\Omega_s}{\widetilde{K}}\left(r^+, \widetilde{K}^+\right) = \sqrt{p}\left(r^+\right)^{\frac{p-1}2}\left(1 + \frac{1}{24}\epsilon^2\left(\widetilde{K}^+\right)^2\right).\label{e:non-reg leading-edge speed}
\end{align}
\ese

\paragraph{Regularized model.}

For the regularized model~\eqref{e:BBM-like PDE}
we solve~\eqref{e:DSW fitting BVP} with $\Omega_0$ given by~\eqref{eq:LDR for BBM}.
This yields
\bse
\label{e:DSW fitting regularized system}
\begin{align}
    \log\left|\frac{\overline{r}}{r^+}\right| &= \frac2{p-1}\left(\frac{\epsilon^2K^2}{48} + \log\left(1 + \frac{\epsilon^2K^2}{24}\right)\right),\label{e:solution to BV1}\\
    \log\left|\frac{\overline{r}}{r^-}\right| &= \frac2{p-1}\left(-\frac{\epsilon^2\widetilde{K}^2}{48} + \log\left|1 - \frac{\epsilon^2\widetilde{K}^2}{24}\right|\right).\label{e:solution to BV2}
\end{align}
\ese
Furthermore, the trailing and leading edge speeds are given by,
\bse
\begin{align}
    s^- &= \frac{\partial \Omega_0}{\partial K}\left(r^-,K^-\right) = \frac{\sqrt{p}\left(r^-\right)^{\frac{p-1}2}\left(1 - \frac{\epsilon^2\left(K^-\right)^2}{24}\right)}{\left(1 + \frac{\epsilon^2\left(K^-\right)^2}{24}\right)^2},\label{e:reg trailing edge speed}\\
    s^+ &= \frac{\widetilde\Omega_s}{\widetilde{K}}\left(r^+,\widetilde{K}^+\right)= \frac{\sqrt{p}\left(r^+\right)^{\frac{p-1}2}}{1 - \frac{\epsilon^2\left(\widetilde{K}^+\right)^2}{24}}.\label{e:reg leading edge speed}
\end{align}
\ese
where again $K^-$ and $\widetilde{K}^+$ are the trailing-edge wavenumber and leading-edge conjugate wavenumbers which can be numerically obtained by solving~\eqref{e:solution to BV1} and \eqref{e:solution to BV2} for $K$ and $\widetilde{K}$, respectively.

\paragraph{Discrete granular lattice.}
We can also apply the DSW fitting to the discrete granular lattice model~\eqref{Granular Crystals}. Modulation equations for a general class of FPUT equations are given in \cite{DHM06}.
Once again, we obtain two boundary value problems, which are similar to~\eqref{e:DSW fitting BVP} but with $K$ being replaced by $k$ (where $K = \epsilon k)$,
\bse
\label{e:DSW lattice BVPs}
\begin{align}
    \frac{dk}{d\overline{r}} &= \frac{\partial\omega_0/\partial\overline{r}}{V\big(\overline{r}\big) - \partial\omega_0/\partial k},\qquad 
    k\left(r^+\right) = 0\label{e:bvp 1 for discrete lattice}\\
    \frac{d\widetilde{k}}{d\overline{r}} &= \frac{\partial\widetilde\Omega_s/\partial\overline{r}}{V\big(\overline{r}\big) - \partial\widetilde\Omega_s/\partial \widetilde{k}},\qquad
    \widetilde{k}\left(r^-\right) = 0,
    \label{e:bvp 2 for discrete lattice}  
\end{align}
\ese
where $\omega_0$ denotes the linear dispersion relation~\eqref{e:linearized DR of DDE}, and $\widetilde\Omega_s$ the conjugate dispersion relation defined as
\begin{equation}
\label{e:conjugate dr for lattice}
    \widetilde\Omega_s\left(\overline{r},\widetilde{k}\right) = -i\omega_0\left(\overline{r},i\widetilde{k}\right) = 2p^{\frac{1}2}\big(\overline{r}\big)^{\frac{p-1}2}\sinh(\widetilde{k}/2),
\end{equation}
with $V(\overline{r}) = \sqrt{p}\,\overline{r}^{\frac{p-1}2}$.
Solving the two boundary value problems in~\eqref{e:DSW lattice BVPs} yields
\begin{equation}
    \label{DDE dsw s_minus} 
    k^- = 4\arc\sec\big[(r^-/r^+)^{\frac{p-1}{4}}\big],
    \quad
    \widetilde{k}^+ = 4\arc\sech\big[(r^+/r^-)^{\frac{p-1}{4}}\big],
\end{equation}
and the associated two edge speeds then read
\bse
\begin{align}
    s^-_{\text{DDE}} &= \frac{\partial\omega_0}{\partial k}\left(r^-, k^-\right) = p^{\frac{1}2}\left(r^-\right)^{\frac{p-1}2}\left(2(r^+/r^-)^{\frac{p-1}2} - 1\right),\label{e:lattice s_minus}\\
    s^+_{\text{DDE}} &= \frac{\widetilde\Omega_s}{\widetilde{k}}\left(r^+,\widetilde{k}^+\right) = \frac{4p^{\frac{1}2}\left(r^+\right)^{\frac{p-1}2}(r^-/r^+)^{\frac{p-1}{4}}\sqrt{(r^-/r^+)^{\frac{p-1}2}-1}}{\widetilde{k}^+}. \label{e:lattice s_plus}
\end{align}
\ese

\section{Numerical validation}
\label{s:numerics}

In this section we report on the results of systematic numerical simulations aimed at verifying the theoretical predictions presented in the previous sections.
Specifically, we numerically measure various features of the dispersive shock waves simulated from the granular lattice DDE and the two continuum models. 
These include the trailing-edge wave number and trailing and leading-edge speeds. 
Moreover, to further understand how the DSWs from each continuum model approximate the one from the granular lattice, we also compare the spatial profile of the DSWs of the two continuum models with that of the granular lattice.

\begin{figure}[b!]
    \centering
    \includegraphics[width=1\linewidth]{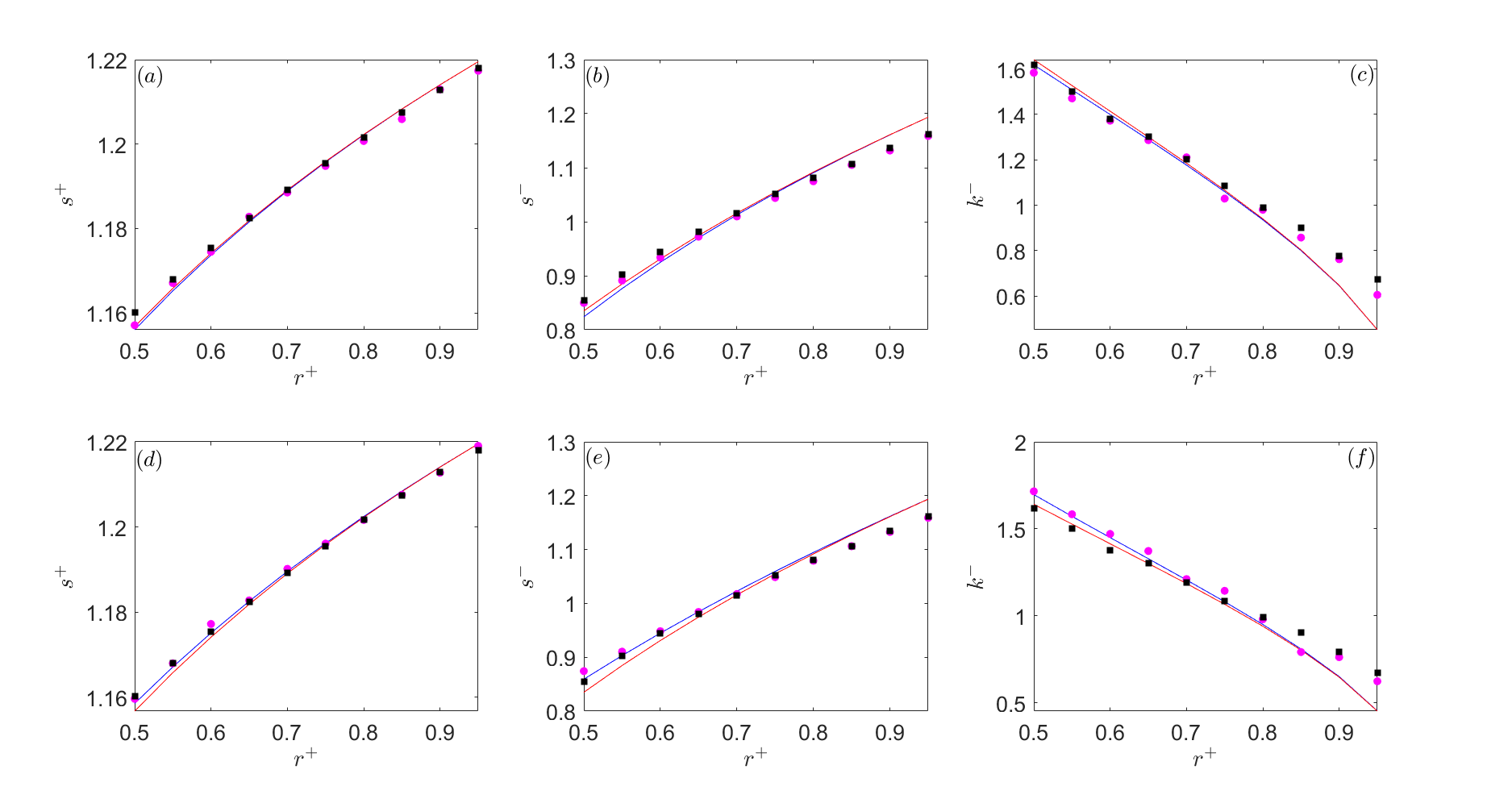}
    \caption{
    Panels~\textbf{(a)}--\textbf{(c)} display the DSW-edge comparisons for the non-regularized model \eqref{e:KdV-like PDE}, while panels \textbf{(d)}--\textbf{(f)} depict the DSW-edge comparisons of the regularized model \eqref{e:BBM-like PDE}. 
    In each panel, the solid red and blue lines refer to the DSW-fitting theoretical predictions on the edge features based on sub-section \ref{s:DSW fitting}, while the magenta circles and black squares depict the numerically measured DSW-edge features of the two continuum models (\eqref{e:KdV-like PDE} and \eqref{e:BBM-like PDE}) and the granular lattice \eqref{Granular Crystals}, respectively. Notice that here $r^- = 1$, and the values of $r^+$ are varied within the region of $[0.5,0.95]$ with a $0.05$ spacing.}
    \label{fig:Non-regularized versus granular lattice}
\end{figure}

\begin{figure}
    \centering
    \includegraphics[width=0.8\linewidth]{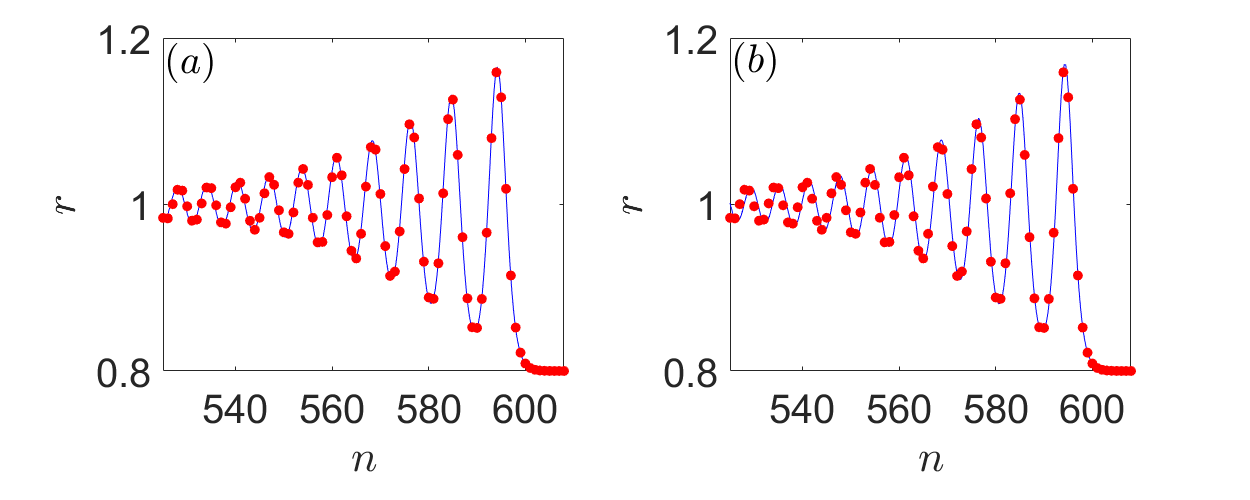}
    \caption{Comparison of the DSW spatial profiles between the continuum models and the granular lattice in the precompression case: The left panel \textbf{(a)} displays the comparison of the DSWs between the non-regularized model \eqref{e:KdV-like PDE} and the associated granular lattice \eqref{Granular Crystals}, while the right panel \textbf{(b)} shows the comparison between the regularized model \eqref{e:BBM-like PDE} DSW with that of the corresponding granular lattice. Notice that the blue solid curves and the discrete red circles in both panels refer to the DSW of the continuum models and the granular lattice, respectively. Both comparisons are shown at $t = 500$ ($T = 50$) with the two backgrounds $r^- = 1$ and $r^+ = 0.8$.
    }
    \label{fig:DSWs comparison in the precompression case}
\end{figure}

\subsection{Precompression case}
\label{s:Lattice with precomp}

We begin with looking at the case when there is precompression in the granular lattice, namely the case when $r^+ > 0$. 
Figure~\ref{fig:Non-regularized versus granular lattice} shows the comparison of the DSW features between the two continuum models and the associated granular lattice. Overall, we can see clearly that the numerically measured DSW edge features agree reasonably well with those from the DSW-fitting theoretical predictions. This also suggests that the spatial profiles of the DSWs of the continuum models and the granular lattice should also agree at a reasonable level. To further confirm this, we compare the DSW spatial profile for the continuum models and the discrete granular lattice which is displayed in Fig. \ref{fig:DSWs comparison in the precompression case}. From this figure, we can see that the red dots, which represent the DSW of the granular lattice, essentially lie on the blue solid curves which are the DSWs of the continuum models. Therefore, through both the DSW-edge features and also the DSW spatial profile comparisons, we conclude that both continuum models \eqref{e:KdV-like PDE} and \eqref{e:BBM-like PDE} provide a good approximatation of the DSWs of the granular lattice when there exists precompression. Note that predictions from the two continuum models proposed here
tend to do better than the KdV approximation, as discussed in more detail in Sec.~\ref{s:crossing regime}.

\subsection{Zero precompression case}
\label{s:Zero precompression}

Next, we switch our focus to the case when there is no precompression in the Riemann initial data, which implies $r^+ = 0$. 
In this case,
the KdV approximation is no longer applicable. The DSW fitting formulas
are also invalid since there is no dispersion around the zero state. Therefore, at this stage, we can only make comparisons from numerical
simulations of the continuum models. While simulations with zero
background posed no difficulty in the regularized PDE model, we did
encounter potential numerical instability in the non-regularized model \eqref{e:KdV-like PDE}. 
The issue seems to stem from small negative values in the numerical solution of the field $r$ in~\eqref{e:KdV-like PDE}. 
To handle this numerical issue, we modify the initial data in~\eqref{e:IC} 
by taking $\delta=1$, so that 
the transition between the two values of the jump is more gradual.
Moreover, we also set $r^+ = 10^{-5}$ 
instead of taking an exact zero lower background value. 
Note, however, that for the regularized model \eqref{e:BBM-like PDE} simulation we still set $r^+ = 0$.

\begin{figure}[t!]
    \centering
    \includegraphics[width=1.05\linewidth]{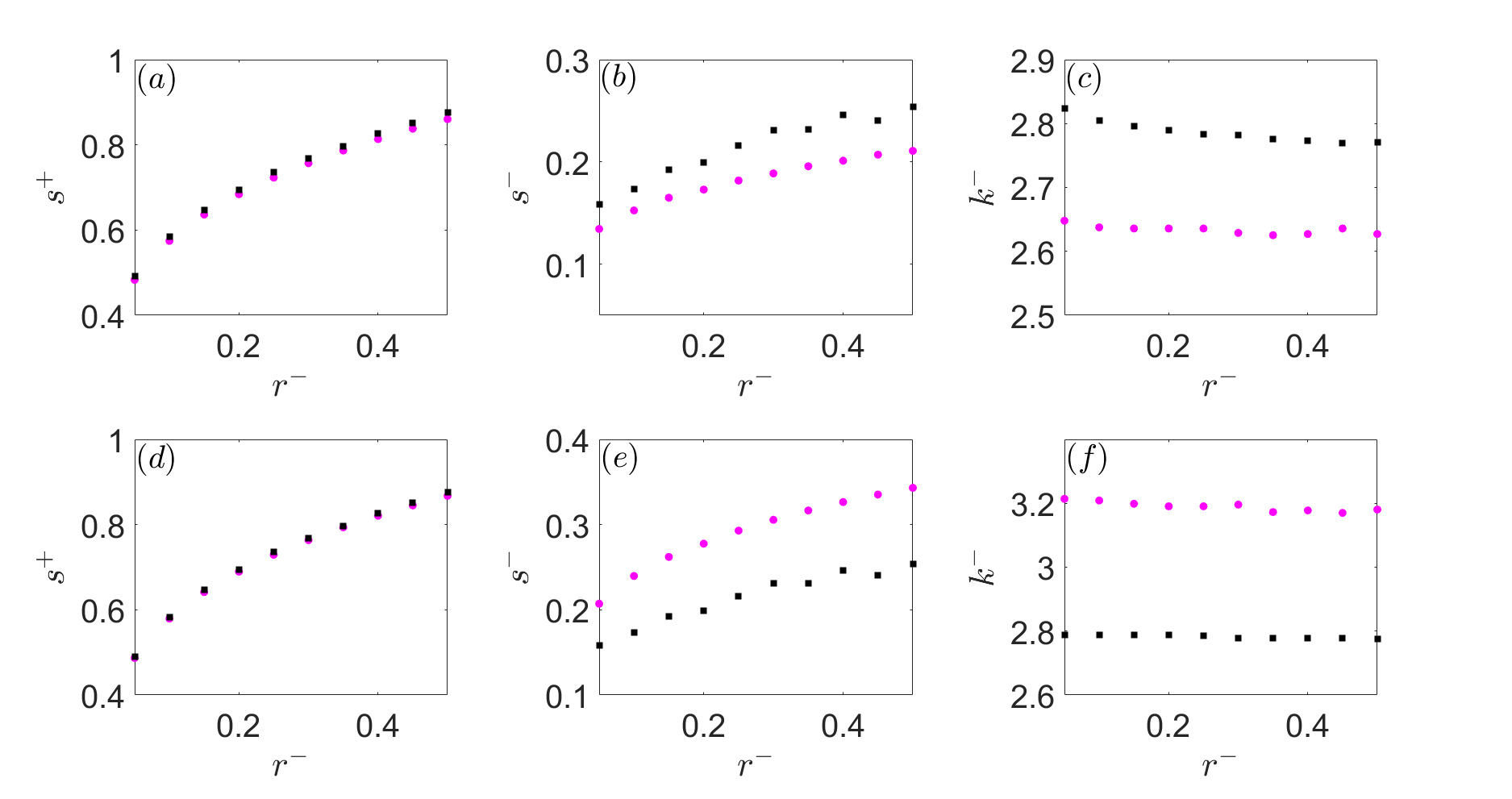}
    \caption{Comparison of the DSW-edge features between the non-regularized model \eqref{e:KdV-like PDE} and the granular lattice \eqref{Granular Crystals} in the zero precompression case $r^+ = 0$. The panels \textbf{(a)}-\textbf{(c)} depict the DSW-edge comparisons of the non-regularized continuum model \eqref{e:KdV-like PDE}. The panels \textbf{(d)}-\textbf{(f)} showcase the DSW-edge comparisons of the regularized continuum model \eqref{e:BBM-like PDE}. Notice that the magenta circles and black squares in each panel above denote the data points of the continuum model and the granular lattice system, respectively.
    }
    \label{fig:DSW-edge feature comp for non-regularized model versus granular lattice}
\end{figure}

\begin{figure}
    \centering
    \includegraphics[width=0.8\linewidth]{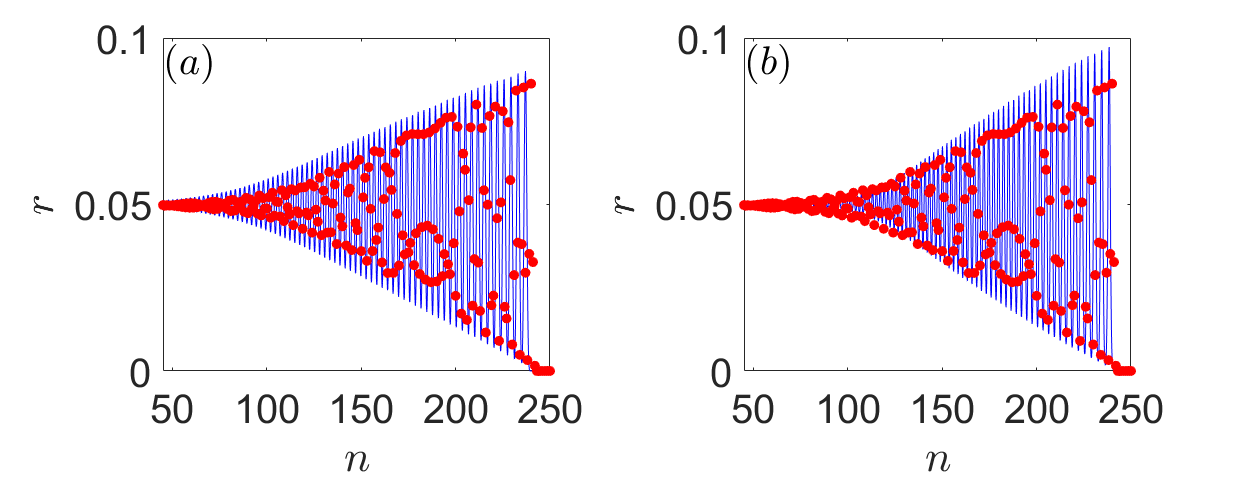}
    \caption{The comparison of the DSWs in the zero precompression case. The panel \textbf{(a)} shows the DSW comparison of the non-regularized model and the associated granular lattice, while the panel \textbf{(b)} depicts the DSW comparison of the regularized model \eqref{e:BBM-like PDE} with the corresponding granular lattice. In both panels, note that the blue solid curves depict the DSWs of the two continuum models\eqref{e:KdV-like PDE} and \eqref{e:BBM-like PDE}, while the red dotis of the DDE \eqref{Granular Crystals}.
    Also notice that both panels depict the evolution dynamics at $T = 50$ ($t = 500$), and the values of all relevant parameters are $r^- = 0.05, r^+ = 0$, and $p = 3/2$.}
    \label{fig:dam_break problem simulations}
\end{figure}

Figure~\ref{fig:DSW-edge feature comp for non-regularized model versus granular lattice} showcases the comparisons of the DSW-edge features between the two continuum models and the granular lattice in the zero precompression case. Based on these comparison results, we can see that the trailing-edge DSW features of both continuum models deviate from those of the granular lattice and hence we are also supposed to expect that the spatial profile comparison of the DSWs shall also deviate at least at the trailing edge of the DSWs. Moreover, as the value of the larger background $r^-$ increases, the solitonic amplitudes of the DSWs also tend to deviate while they agree nicely when there is a small jump (i.e., $r^- = 0.05$). 
Finally, Fig.~\ref{fig:dam_break problem simulations} displays the spatial profile comparisons of the DSWs between the two continuum models and the granular lattice. From the left panel which shows the DSW comparison between the non-regularized model \eqref{e:KdV-like PDE} and the granular lattice \eqref{Granular Crystals}, we see that the non-regularized model DSW tends to have a better agreement on the leading edge with the lattice DSW, while from the right panel, the regularized model DSW has a better agreement on the trailing edge. 

\subsection{Transitioning from finite precompression towards no precompression}\label{s:crossing regime}

\begin{figure}[t!]
    \centering
    \includegraphics[width=1.05\linewidth]{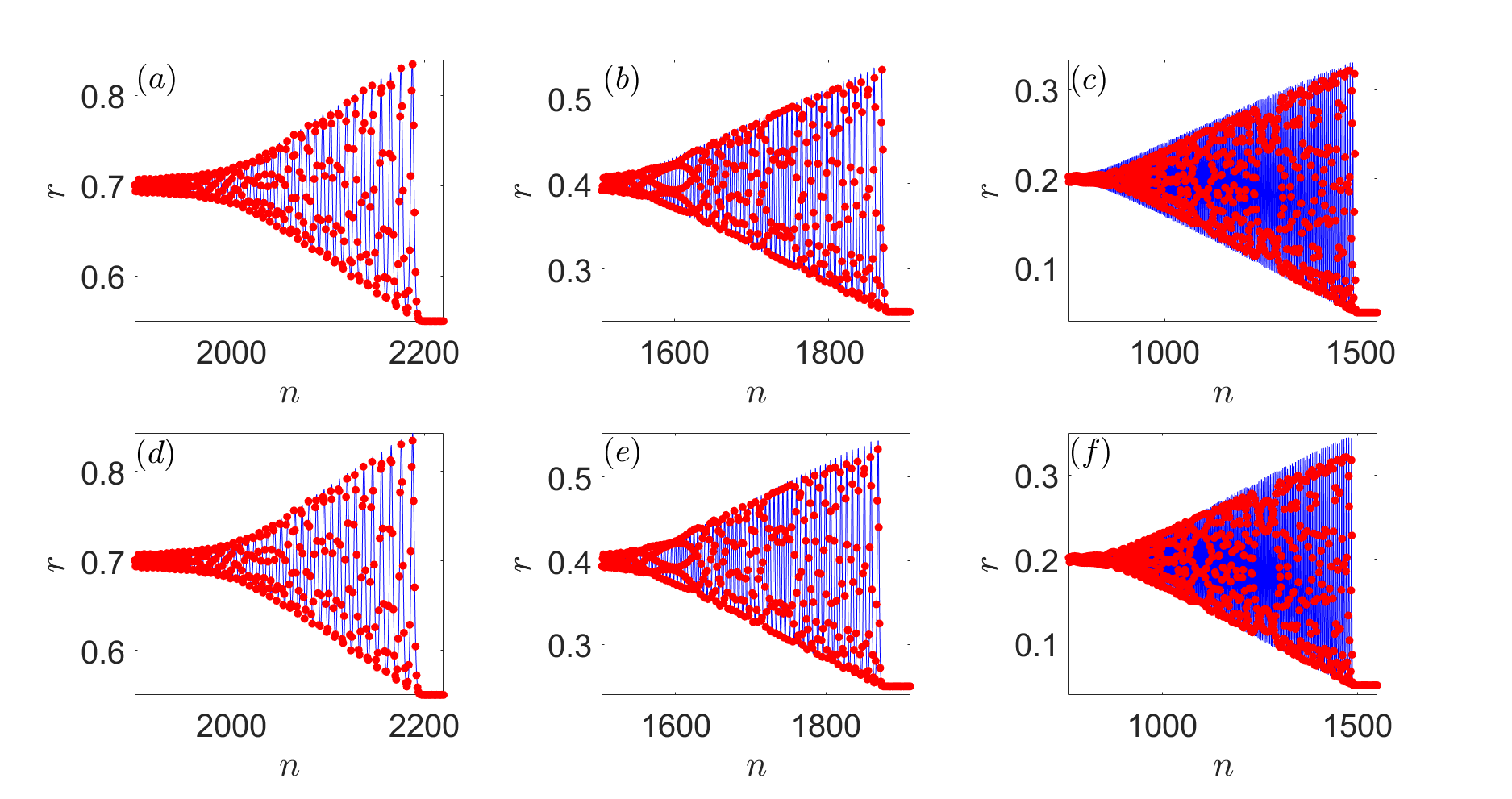}
    \caption{The DSW spatial profile comparison between the two continuum models (\eqref{e:KdV-like PDE} and \eqref{e:BBM-like PDE}) and the granular chain \eqref{Granular Crystals} at $t = 2000$ ($T = 200$). Panels \textbf{(a)}-\textbf{(c)} depict the DSW comparison of the non-regularized model \eqref{e:KdV-like PDE} and the granular chain \eqref{Granular Crystals}, while panels \textbf{(d)}-\textbf{(f)} display the DSW comparison of the regularized model \eqref{e:BBM-like PDE} and the granular lattice \eqref{Granular Crystals}. Notice that the values of smaller backgrounds $r^+$ are $0.55, 0.25, 0.05$ from the leftmost to the rightmost panels above, while the jump in the initial conditions is always fixed $\Delta = 0.15$. Moreover, note that the blue solid curve and the discrete red dots depict the DSW of the two continuum models and the discrete granular chain, respectively.
    } 
    \label{fig:Non-reg DSWs fixing the jump = 0.15}
\end{figure}

One natural question that may arise in light of the results of the previous section is the following: since we have seen a good agreement of the DSWs in the finite precompression case from section \ref{s:Lattice with precomp} and a worse agreement in the zero precompression case, then one may ask how the comparison transitions from one case to the other. 
To address this question,
we run further simulations of both the continuum and the discrete granular models with a fixed jump, denoted by $\Delta$, in the Riemann initial data to be $\Delta = 0.15$ and with the values of the smaller background $r^+ = [0.05, 0.25, 0.55]$ (so that $r^- = [0.2, 0.4, 0.7]$) and then perform the spatial profile comparison of the DSWs just as in the previous two sub-sections \ref{s:Zero precompression} and \ref{s:Lattice with precomp}. 
Figure~\ref{fig:Non-reg DSWs fixing the jump = 0.15} shows all relevant comparisons of the DSWs between the two continuum models and the granular chain. We observe clearly that in the comparisons, there is a trend that larger values of the background $r^+$ gradually lead to a better agreement between the continuum model DSW and that of the discrete granular lattice.
We emphasize, however, that the agreement still remains quite good in the limit of no precompression, especially when compared with the situation for the KdV approximation (compare Fig.~\ref{fig:KdV DSW spatial profile comparisons} in Section~\ref{s:KdV approx} with Fig.~\ref{fig:Non-reg DSWs fixing the jump = 0.15}).

\section{Conclusions and future challenges}

In the present work, we have developed two models
that represent suitable {\it unidirectional} 
continuum limits of
the granular crystal setting in the presence, and
{\it even in the absence} of precompression. 
One of these models was a generalized form of
the KdV equation, while the other one was a regularized
form thereof, in a way reminiscent of the
derivation of the BBM equation. These models
were inspired by the analogy of the case where
linear dispersion exists with KdV and the usage
of the latter to prove detailed existence and
stability results for FPUT-type systems that can be reduced
to the KdV in a suitable long-wavelength limit. A natural
hope is that similar developments could arise in 
the context of the models presented herein. This poses
an interesting challenge for further rigorous mathematical analysis.

As a starting point towards appreciating the
potential usefulness of these models, the 
analysis herein focuses on aspects related
to dispersive shock waves (DSWs). More specifically, 
we explored the traveling wave and periodic wave
aspects of the models, as the former emerge at
the front of DSWs and the latter are the (self-similarly)
modulated waveforms that constitute the DSWs. The
conservation laws of the models were obtained
en route to leveraging them in order to derive
the Whitham modulation equations for the proposed
models. Given the complexity of the latter, as
is commonly done, we obtained both special case (such
as the rarefaction wave) and asymptotic results,
such as those developed by the well-established
DSW fitting method 
(although the latter is far less widespread in
spatially discrete settings). The 
findings both for the leading and for the 
trailing edge of the DSWs were compared to the
original discrete model and, where appropriate
(i.e., when precompression was present) to the
KdV findings. The latter was an especially
important comparison as it showcased that when
the precompression exists but is ``weak''
(i.e., close to the sonic vacuum limit), the
KdV equation is still not an adequate approximation of
the granular chain. Instead, the newly proposed
models are far more accurate in their approximation
of the discrete setting, rendering their usage
in this limit a suitable ``intermediate level''
tool for the long-wavelength description of the
discrete model. Importantly, the comparison
of the developed models with the discrete case
is reasonable even near the sonic vacuum limit.

Naturally, this study and the models it proposed
suggest a number of interesting questions for future
study. From an analysis perspective, a detailed 
understanding of the well-posedness properties
of the present models is of particular interest. 
Similarly, the rendering of the connection
of the granular chain with the present models
more rigorous, by analogy to the FPUT in connection
with the KdV, would also be very valuable towards
using the PDE analysis to obtain a systematic lattice
understanding. From the perspective of the DSW
questions raised herein, arguably, the most
pressing one concerns the properties (e.g., 
strict hyperbolicity and genuine nonlinearity~\cite{El2005,Hoefer2014}) of the
Whitham modulation equations. A further analysis of the
Whitham equations to appreciate features of
the DSW appears to us to be a central theme
of emerging interest in lattice dispersive hydrodynamics.
Finally, there are numerous 
motivations ~\cite{magBreathers2D,Leonard11,l11,andrea}
towards the study of 2D lattice problems, yet it seems
that lattice explorations at that level are very limited.
Such a direction is particularly worthwhile of exploration
and arguably the models herein pave the way towards the
potential development of Kadomtsev-Petviashvili~\cite{ablowitz2011nonlinear} continuum
analogues of such lattice settings. These directions
are currently under investigation and relevant findings
will be reported in future publications.

\setcounter{equation}0
\def\thesubsection{A.\arabic{subsection}}
\def\theequation{A.\arabic{equation}}
\addcontentsline{toc}{section}{Appendix}

\section*{Appendix: Calculation of trailing edge speeds and leading edge speeds}
\label{a:speeds}

In order to test the theoretical predictions for the features of the DSW, one must validate them against the results of direct numerical simulations.
In this appendix, we discuss the relevant methods utilized to numerically measure the edge features of the DSWs of the two continuum models and the granular lattice. 

For the leading-edge speed, denoted by $s^+$, we first treat the highest peak of the dispersive shock wave as its associated leading edge location. We then keep track of the $x$ coordinates of the leading edge for multiple time snapshots $\{t_i\}$, and finally we compute numerically the slope of the line constructed by the $x$ locations of the leading edge and the associated time snapshots, and then treat the slope as the speed of the leading edge, $s^+$.

On the other hand, for the trailing-edge features which include both the trailing-edge wavenumber $K^-$ and speed $s^-$, we first define the following two quantities,
\begin{align}
    a^{u} = r^- + \frac{\left|r^- - r^+\right|}{N} ,\qquad
    a^{l} = r^- - \frac{\left|r^- - r^+\right|}{N},
\end{align}
where $N$ is a positive integer.
Then we utilize all the local maxima and minima of the dispersive shock wave which fall within the following two interval windows, respectively, 
\begin{align}
    I^{u} = \big(a^{u} - \nu, a^{u} + \nu\big),\qquad
    I^{l} = \big(a^{l} - \nu, a^{l} + \nu\big),
\end{align}
where $\nu \in \mathbb{R}$ is a number determining the width of the two windows. 
On the one hand, we fit these local maximum peaks of the dispersive shock wave with a line. Similarly, we fit a line through all the local minima and then treat their intersection as the location of the trailing edge. To compute the trailing edge speed, we simply use the distance traveled by the trailing-edge and divide it by the total simulation time. Mathematically, if we denote the trailing-edge speed by $s^-$, then it is simply,
\begin{equation}\label{e:trailing-edge speed formula}
    s^- = \frac{X_--X_0}{T_f},
\end{equation}
where $X_-$ and $X_0$ refer to the trailing-edge location of the DSW at the final-time snapshot $T = T_f$ and at the initial time $T = 0$, and $T_f$ denotes the total simulation time. Finally, for the wavenumber of the trailing edge, we first note that we can relate the wavenumber of the discrete granular system \eqref{Granular Crystals} with that of the continuum model through the following relation,
\begin{equation}\label{e:relation of wavenumbers}
    K = k/\epsilon,
\end{equation}
and we will compare the trailing-edge wavenumber at the level of the lattice (i.e., at the level of $k$). Finally, we discuss also the method utilized to measure the trailing-edge wavenumber of the DSWs of the KdV model \eqref{e:KdV reduction} and the granular lattice \eqref{Granular Crystals}. On one hand, for the trailing-edge wavenumber measurement of the DSW of the KdV model, we find the two adjacent local peaks of the DSW right next to the trailing-edge location $X_-$ and then record the associated $x$ coordinates of such two local peaks and we denote them by $X_{1}$ and $X_2$, then the trailing-edge wavenumber of the KdV DSW is simply calculated as follows,
\begin{equation}\label{e:Regular KdV k_minus measurement}
    k^- = \frac{2\pi\epsilon}{X_2-X_{1}}.
\end{equation}
For the trailing-edge wavenumber of the granular chain, since the spatial resolution may not be sufficiently good to use the same approach in measuring the wavenumber of the continuum KdV model, we instead utilize a temporal approach which can yield more accurate measurement of the wavenumber of the DSW of the discrete granular lattice. The way to numerically compute the trailing-edge wavenumber is as follows: We first write the solution as
\begin{equation}
\label{e:solution definition}
    u(n,t) = f(kn - \omega t) = f\bigg(\omega\left(\frac{k}\omega n - t\right)\bigg),
\end{equation}
and then we first compute the value of $\omega$ as follows
\begin{equation}\label{e:value of omega}
    \omega = \frac{2\pi}{t_2 - t_1},
\end{equation}
where $t_1$ and $t_2$ are two adjacent time snapshots of the time-series data of $u(n^-,t)$ where $n^-$ denotes the trailing-edge location of the granular lattice DSW computed by the method discussed before. Then, we look at the time-series data of $u(n^-+1,t) = f\big(\omega\big((k/\omega)\,n^- - (t - k/\omega)\big)\big)$, 
so that the value of ${k}/\Omega$ can be measured by examining the horizontal $t$-axis distance between the two time-series data of $u(n^-,t)$ and $u(n^-+1,t)$. Lastly, multiplying the values of $\omega$ measured in Eq.~\eqref{e:value of omega} with ${k}/\Omega$ yields the value of the trailing-edge wavenumber of the discrete granular lattice DSW.

\bibliographystyle{unsrt}

\bibliography{main}

\end{document}